\documentclass{raa}            
\pdfoutput=1

\usepackage{graphicx,times}             
\usepackage{natbib}
\usepackage{amssymb,amsmath}
\bibpunct{(}{)}{;}{a}{}{,}
\usepackage{adjustbox}

\graphicspath{{./plots/}}
\newcommand{\LCDM}{$\Lambda$CDM}
\newcommand{\ri}{\mathrm{i}}
\newcommand{\rp}{\mathrm{p}}
\newcommand{\E}{\mathcal{E}}

\newcommand{\ud}{\mathrm{d}}
\newcommand{\mpch}{\mathrm{Mpc}\,h^{-1}}
\newcommand{\msunh}{\mathrm{M}_\odot\,h^{-1}}
\newcommand{\msun}{\mathrm{M}_\odot}
\newcommand{\deltac}{\delta_{\rm sc}^{\rm lin}}
\newcommand{\spheric}{\textsc{SpheriC}}

\usepackage[pagebackref=true]{hyperref}

\begin{document}

  \title{The many boundaries of the stratified dark matter halo}

   \volnopage{Vol.0 (20xx) No.0, 000--000}      
   \setcounter{page}{1}          

   \author{Jiaxin Han 
      \inst{1,2,3}
   }

   \institute{Department of Astronomy, School of Physics and Astronomy, Shanghai Jiao Tong University, Shanghai 200240, China; {\it jiaxin.han@sjtu.edu.cn}\\
    \and 
    State Key Laboratory of Dark Matter Physics, Shanghai Jiao Tong University, Shanghai 200240, China\\
    \and 
    Key Laboratory for Particle Astrophysics and Cosmology (MOE) \& Shanghai Key Laboratory for Particle Physics and Cosmology, Shanghai Jiao Tong University, Shanghai 200240, China\\
\vs\no
   {\small Received 20xx month day; accepted 20xx month day}}

\abstract{
We review the physics of halo collapse giving rise to various halo boundaries, as well as their identification, observation, and applications. The classical halo is typically defined as a monolithic, virialized object enclosed within its virial radius -- a definition which, however, does not account for ongoing halo growth. Continuous accretion causes the orbits of infalling particles to shrink over time, confining newly accreted material in a growing layer outside the virialized region. Several novel halo boundaries, such as the splashback and depletion radii, have recently been proposed to characterize this growth layer from different perspectives. Along with the turnaround radius, which operates on an even larger scale to enclose the entire infall region, these multiple boundaries comprise an extended view of a dark matter halo as a stratified structure. Theoretical models can largely explain the existence of various boundaries, while challenges remain in providing unified and quantitative predictions of their properties. The multiple boundaries open new avenues for observing halo growth and may substantially improve our understanding and modeling of cosmic structure formation. We provide a python package, \textsc{SpheriC}, implementing the key spherical collapse models.
\keywords{cosmology: dark matter; large-scale structure of Universe; galaxies: haloes}
}

   \authorrunning{J. Han}            
   \titlerunning{Halo boundaries }  

   \maketitle

%
%
\section{Introduction: the requirement for a physical halo}           
\label{sect:intro}
In the 1970s, evidence became strong that galaxies are surrounded by an extended halo component of invisible matter~\citep[e.g.,][]{Rood72,Ostriker74,Rubin78,Rubin80}. Theorists started to consider the formation of such a halo by modeling how collisionless matter could be accumulated around a galaxy component, in a cosmological context, in order to explain the observed galaxy rotation curve as well as the profile of elliptical galaxies and the halo of X-ray clusters~\citep[e.g.,][]{GunnGott72, Gott75, Gunn77}. With the advancement of numerical simulations, the formation of halos from a perturbed primordial density field can now be well resolved, revealing a highly universal internal structure across mass scales in the cold dark matter paradigm~\citep[e.g.,][]{nfw96,nfw97,Wangjie20}. These ubiquitous and universal dark matter structures, commonly found around galaxies and galaxy clusters and approximately in equilibrium at least in the inner part, are known as dark matter halos.

Dark matter halos are nowadays fundamental units for the understanding of cosmological structure formation and galaxy formation. 
In the halo model of the largescale structure~\citep[see][for a review]{CooraySheth02}, it is essentially assumed that all the mass in the Universe can be partitioned into haloes of different sizes. 
The effectiveness of this approach relies on the condition that the statistical properties of these haloes, including their internal structure, sizes, and spatial distribution, are relatively easy to describe. To achieve this, it is imperative that a dark matter halo is defined as a physical object rather than any arbitrary mathematical construction. 
The physical laws shaping the halo then help to maintain universal and understandable properties of these objects. 

The demand for a physical halo can be stronger when using halos to model galaxy formation. Due to its collisionless nature, dark matter condensate to form a halo prior to the condensation of baryons. The dark matter halo then provides the environment within which galaxies form and evolve~\citep{WhiteRees78, MoBook}. A good physical model can reveal key physical properties that regulate the baryonic processes, including the accretion, stripping, recycling, cooling, heating, condensation,  ejection, rotation, and virialization of various baryonic components happening inside and around the halo.

The conventional halo definition aims to identify a virialized object, with its boundary--the virial radius--defined where the mean enclosed density reaches a characteristic threshold.
This definition is motivated by the monolithic spherical collapse model, and is able to largely depict an approximately virialized region of the halo. Besides, the virial radius is relatively easy to calculate from the particle distributions in simulations or from the density profile measured in observations using gravitational lensing or dynamics. As such, it has found wide applications in both theories and observations. Despite this, the virial-radius-based definition of a halo suffers from a few fundamental limitations. Below, we highlight three major limitations.

\begin{itemize}
    \item \textbf{Pseudo evolution}. The characteristic density within the virial radius is usually defined (or rather assumed) to be a characteristic factor times the background density of the Universe. For low mass halos that have largely stopped mass accretion, their density profiles are approximately frozen~\citep[e.g.,][]{Cuesta08,Zemp14}. However, the evolution of the background density means the virial radii identified around these halos still evolve over time. This size evolution no longer reflects the physical evolution of the halo, and is thus referred to as pseudo-evolution~\citep{PseudoEvo}. It reveals a limitation of the virial radius in providing a \textit{physical} description of the halo size, especially for those located in a strong tidal environment that can inhibit halo growth~\citep{Huiyuan07}. Note that in an ideal universe as assumed in the spherical collapse model, the halo profile is expected to co-evolve with the background density, so the pseudo-evolution is ultimately reflecting a deviation of the actual profile evolution from the theoretical expectation.
    \item \textbf{Halo exclusion}. In order to use halos as fundamental units to reconstruct the cosmic density field, it is necessary that halos are tightly packed around each other and fill up the entire space. However, in reality, halos bounded by the virial radius do not satisfy this constraint, as there is always non-virialized material around a halo outside the virial radius. This leads to space and mass not covered by halos when building a halo model for the largescale structure, making the halo model \textit{inaccurate} in describing the matter distribution on the inter-halo scale. This problem is commonly known as the halo exclusion issue, with many efforts in the literature to make up for the ambiguity in modeling the mass distribution on the inter-halo scale~\citep[e.g.,][]{tinker2005mass,HayashiWhite2008,van2013cosmological,Garcia19,Garcia21,ADM1,ADM2,ZhouHan23,ZhouHan25}.
    
    \item \textbf{Assembly bias}. On the largest scale, the spatial distribution of halos can be specified by the halo bias, which describes the largescale clustering of halos relative to the clustering of matter. The mass dependence of halo bias is well understood in classical theories of halo clustering by considering the conditional halo mass function in different environments~\citep[e.g.,][]{BBKS,CK89,MW96}. However, simulations show that halo bias also depends on a number of other halo properties in addition to mass~\citep[e.g.,][]{Gao05,Gao07,Jing07,Bett07}. In particular, recent works highlight that the environment outside the virial radius of a halo plays a key role in determining the halo bias~\citep{Han19,TidalAni19}. This means that the virial radius, including the mass distribution inside it, is \textit{incomplete} for describing the physical property of a halo on the largest scale.
\end{itemize}

These issues occur on the one-halo, two-halo and many-halo scales respectively. All of them are associated with the fact that the virial radius of a halo only describes the relatively inner part of a halo, while a realistic halo is embedded in, and interacting with, the largescale environment. To overcome this limitation, recent works start to incorporate the surrounding non-virialized envelopes into the domain of a halo. 
Consequently, a complete and modern view of a halo describes it as a stratified object characterized by multiple boundaries. 

\begin{figure}
\centering
    {\includegraphics[width=0.5\textwidth]{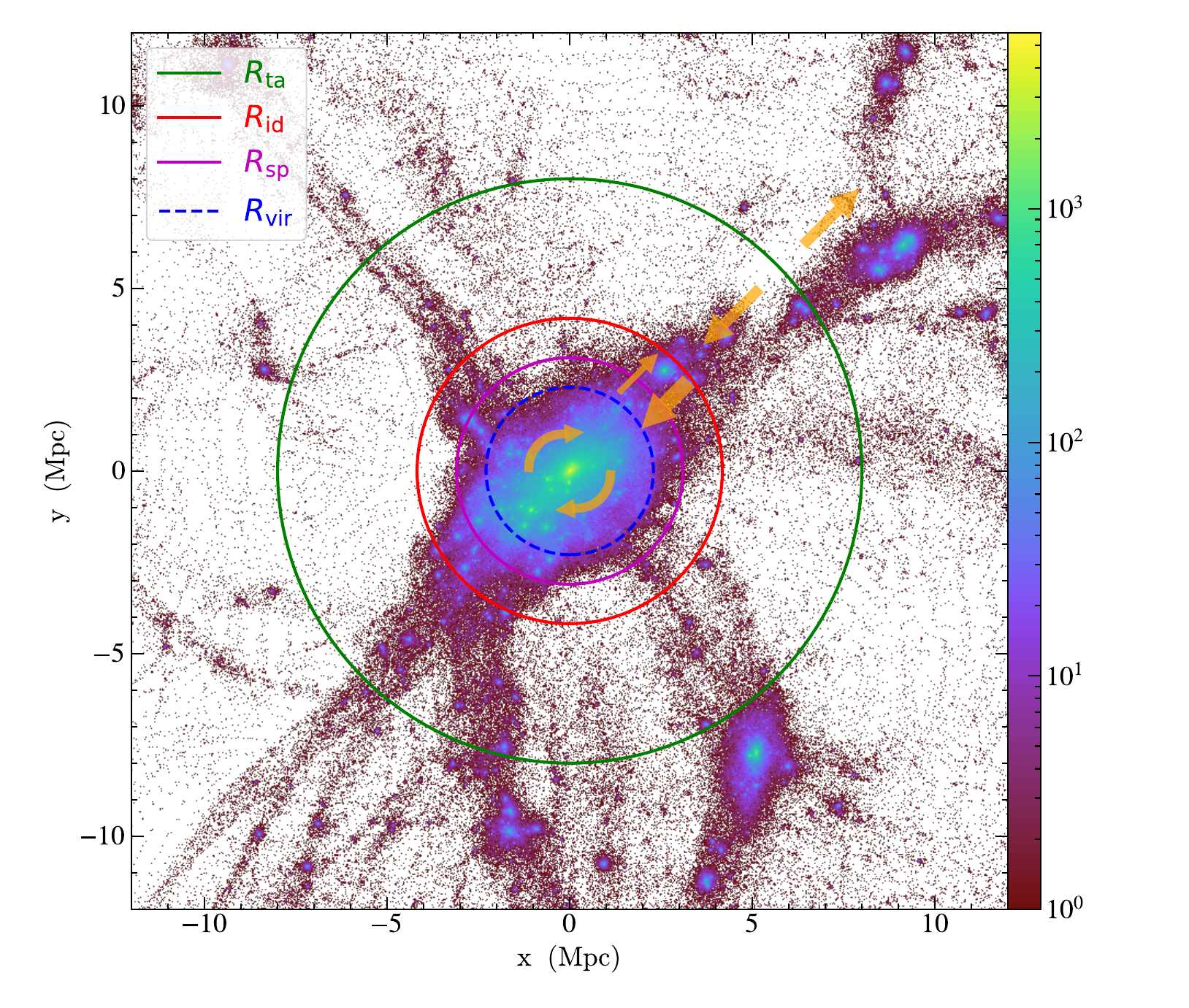}}%
    \raisebox{-11pt}[0pt][0pt]{\includegraphics[width=0.49\textwidth]{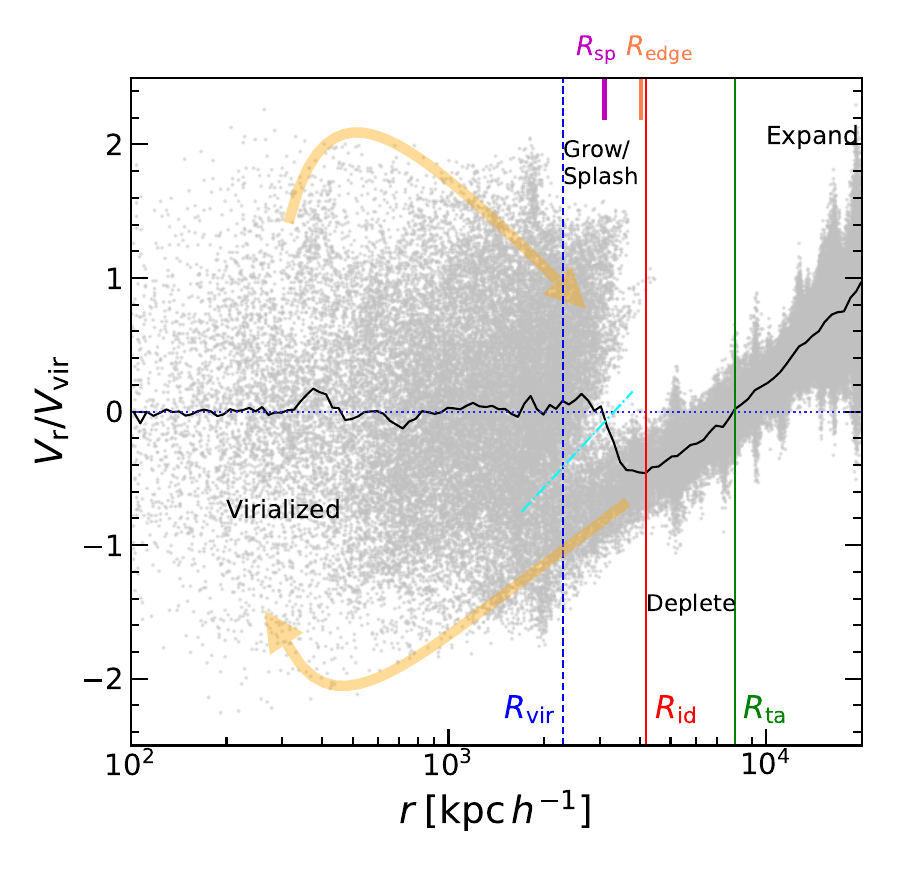}}
    \caption{The layers and boundaries around a cluster-size dark matter halo at $z=0$ from a cosmological $N$-body simulation. \textit{Left:} Projected dark matter distribution in a $2\mpch$ slice along the $z$ direction. The color-bar indicates the projected particle count in each pixel. \textit{Right:} The dark matter particle distribution in phase space (halo-centric radius versus radial velocity) around the same halo. The black solid curve shows the average radial velocity profile. The vertical colored lines mark the virial ($R_{\rm vir}$), depletion ($R_{\rm id}$), and turnaround radius ($R_{\rm ta}$), respectively. These boundaries divide the space around the halo into the virialized, growing, depleting and expanding regions. The short solid lines on the top axis mark the locations of the splashback radius, $R_{\rm sp}$, and the edge radius, $R_{\rm edge}$. The cyan dotted line illustrates the approximate separation between the orbiting and infalling particles. Some of these radii are also shown in the left panel. In both panels, the orange arrows indicate the mass flow pattern.}\label{fig:halo_region}
\end{figure}

Figure~\ref{fig:halo_region} shows the major layers and the corresponding boundaries in dark matter around an example halo in cosmological simulations, including: 
\begin{itemize}
    \item \textbf{Virial radius:} In the innermost part, the radial velocities are mostly symmetrically distributed, with an average radial velocity close to zero, such that there is no net inflow or outflow and the mass distribution stays stable. This region corresponds to the virialized classical halo characterized by the virial radius. 
    \item \textbf{Turnaround radius:} On the largest scale, particles follow the Hubble flow with an increasing velocity over distance. They start to decouple from the Hubble flow and fall towards the halo at the {turnaround radius} where the radial velocity is zero. 
    \item \textbf{Splashback radius:} Within the turnaround radius, material falls towards the halo with an accelerating velocity, passing through the pericenter in the inner halo and moving outward again. During this process, the halo is also growing in its mass, so that the gravitational pull experienced by the particles when they move out is stronger than that when they fell in at an earlier time, resulting in a reduction in subsequent apocenter distance over time. This process is known as the splashback process, which confines the particles to a smaller region than the original turnaround domain. As a result, the density profile around the halo also drops steeply in this region, and a corresponding {splashback radius} is usually defined at the steepest slope location~\citep{DK14,Adhikari14,More15}. 
    \item \textbf{Depletion radius:} The existence of the outflowing particles in the splashback region also causes the average radial velocity of the particles to brake before eventually approaching zero in the virialized region, as shown by the black solid curve in the right panel. The net mass flow rate thus first accelerates and then decelerates towards the halo. Due to continuity, the density within the accelerated infall region drops over time, while that in the decelerated infall region grows. The transition between the two defines the {depletion radius}~\citep{FH21,Gao23,ZhouHan23}. 
    \item \textbf{Truncation/edge radius:} According to whether a particle has already passed its first pericenter passage, the halo can also be decomposed in phase space into an infalling and an orbiting component~\citep[e.g.,][]{Tomooka20, Aung21,DiemerDecompose,Garcia23}. In fitting the orbiting density profile, a truncation radius is often introduced to describe the location where this density falls off steeply~\citep{DiemerDecompose2,DiemerDecompose3}. The truncation and splashback radii are intimately related since they both delineate the outermost orbits, but the truncation radius does not necessarily occur at the steepest slope of the density profile. The depletion radius is located very close to the so called {edge radius} defined in \citet{Aung21}, where the fraction of orbiting or splashback particles drops to a negligible fraction ($1\%$) of the total particles at the same radius.
\end{itemize}
More comprehensive introductions to various boundaries can be found in section~\ref{sec:boundary}.

Theoretical models characterizing the different boundaries can be traced back to when the concept of a halo first emerged.  
The pioneering work of \citet{GunnGott72} realized that the matter distribution in an extended area around a primordial perturbation can be bound to the perturbation and collapse to provide the halo material. Coupled with the virialization argument and generalized to different cosmologies in many subsequent works, these are now known as the spherical collapse (SC) model.
\citet{Gott75} assumed that the accreted material can be approximately modelled as mostly staying at their apapsis, giving rise to a mass profile consistent with that of elliptical galaxies. \citep{Gunn77} further studied the virialization of the collapsing shells under the assumption of a self-similar mass profile, which allows the shells to decay in the apocenter distance while maintaining the mass profile prediction. Self-similarity was further exploited in \citet{FG84} and \citet{Bertschinger85} to compute the self-consistent mass profiles and orbits in the Einstein-de-Sitter (EdS) universe, for both collisionless and collisional components. These models are often referred to as the secondary infall or self-similar models of spherical collapse, and provide key insights on the formation of the halo profile and the layers of a halo. Note that secondary infall is already studied in \citet{GunnGott72}.




Despite the wide applications of these models and a number of recent developments~\citep[e.g.,][]{Adhikari14,Shi16b,Shi16}, a self-contained and coherent review of them is largely missing. This review intends to fill this gap, by covering both the classical spherical infall and the self-similar secondary infall models which provide the physical foundation for the various halo boundaries.  
We also briefly review recent simulation analysis, theoretical applications, and observational measurements of the new boundaries. A heuristic and insightful introduction to the models of self-similar spherical collapse can be found in the lecture notes by \cite{ShiNotes}. The current review, however, will take a more rigorous and hands-on approach in deriving the various equations. 
A more observation-oriented review on the outskirts of galaxy clusters can be found in \citet{ClusterBoundaryRev}.

Beyond reviewing the classical models, this work also provides some original insights and extensions to the theories to bridge the gaps across the literature. In particular, we provide a derivation of the recently proposed depletion radius in the self-similar collapse model for the first time (section~\ref{sec:ss_solution}).
A python package, \textsc{SpheriC}, implementing the monolithic and self-similar collapse models is presented in Appendix~\ref{app:spheric}.

\section{Classical boundaries from monolithic spherical collapse}
\label{sect:theory}
The classical definition of a dark matter halo has been primarily based on the virial radius, which is the expected final radius of a virialized halo in the spherical collapse model. \citet{GunnGott72} first calculated analytically the collapse process of a uniform spherical region embedded in a background universe. Such a region could represent a perturbation that survived after recombination, which provides the initial condition for a dark matter halo. They considered such a perturbation to be initially co-expanding with the background universe. The extra gravity from the perturbation then causes this region as well as its surroundings to expand at a different pace from the background universe and may eventually turnaround and collapse. 

Below we provide a more accessible derivation of the model, under the $\Lambda$ Cold Dark Matter (\LCDM) cosmology. In absence of dark energy, the evolutions of the sizes of both the perturbed region and the background universe are purely gravitational processes, sharing similar scaling laws for the dynamical timescale except for differing coefficients. This results in a key prediction of the model that \emph{the characteristic densities of halos are universally in proportion to the background density}.

\subsection{Dynamical Equations}
Consider a spherically symmetric mass distribution with a radius $r$ and an enclosed mass $M$. Below we will also refer to this region as the perturbed region. The equation of motion for the shell of radius $r$ can be written as
\begin{equation}
 \ddot{r}=-\frac{GM}{r^2}+\frac{\Lambda}{3}r,\label{eq:SC_Acc}
\end{equation} where the cosmological constant is providing a repulsive force. Integrating this equation leads to the energy conservation equation,
\begin{equation}
 \frac{1}{2}\dot{r}^2-\frac{GM}{r}-\frac{\Lambda r^2}{6}=\E, \label{eq:Newton_E}\\
\end{equation}
where $\E$ is the specific energy of the shell. Note the above equations apply to not only uniform regions but any spherically symmetric mass distributions. 

Rewriting the above equation in terms of the enclosed density, $\rho=M/(\frac{4\pi }{3}r^3)$, we get 
\begin{align}
 \frac{\ddot{r}}{r}&=-\frac{4\pi}{3}G\rho+\frac{\Lambda}{3},\label{eq:Newton_Acc}\\
\left(\frac{\dot{r}}{r}\right)^2&=\frac{8\pi G}{3}\rho+\frac{\Lambda}{3}+\frac{2\E}{r^2}.\label{eq:Newton_Vel}
\end{align} 

Now it is straightforward to recognize these equations as a local version of the Friedmann equations governing the expansion of the background universe,
\begin{align}
 \frac{\ddot{R}}{R}&=-\frac{4\pi}{3}G\bar{\rho}+\frac{\Lambda}{3},\label{eq:Friedmann_Acc}\\
 \left(\frac{\dot{R}}{R}\right)^2&=\frac{8\pi G}{3}\bar{\rho}+\frac{\Lambda}{3}-\frac{kc^2}{R^2},\label{eq:Friedmann_vel}
\end{align} 
where $R$ is the (dimensional) scale factor, $\bar{\rho}$ is the mean density of the universe, and $k$ is the curvature parameter. In Equation~\eqref{eq:Friedmann_vel} we have also ignored any radiation component in the matter dominated epoch. When the density of the spherical region equals the background density, Equations~\eqref{eq:Newton_Acc} and \eqref{eq:Friedmann_Acc} become identical. However, the spherical region can still expand differently from the background universe depending on its energy, $\E$, which is in turn determined by its initial conditions.

\subsection{Analytical solution in absence of dark energy}\label{sec:sc_sol_eds}
For $\Lambda=0$, a simple analytical solution exists for Equation~\eqref{eq:Newton_Acc}. This solution also applies to the $\Lambda$CDM model at a sufficiently early time, where $\Omega_\Lambda\ll \Omega_{\rm m}$ so that the $\Lambda$ term can be ignored. 

We are interested in a bound system with $\E<0$, which would eventually stop expanding and collapse back at a maximum radius of 
\begin{equation}
r_{\rm ta}=\frac{GM}{-\E}.\label{eq:rta}
\end{equation}
Integrating both sides of $\ud t=\ud r/\dot{r}$ with the help of Equation~\eqref{eq:Newton_Vel}, the solution can be found in a parametric form as
\begin{align}
	r&=r_{\rm ta} \frac{1-\cos\phi}{2},\label{eq:sc_sol_r}\\
	t-t_0&=t_{\rm ta} \frac{\phi-\sin\phi}{\pi},\label{eq:sc_sol_t}
\end{align} where $\phi\in[0, 2\pi]$, and
\begin{equation}
t_{\rm ta}=\frac{\pi}{2}\frac{r_{\rm ta}}{\sqrt{-2\mathcal{E}}}=\sqrt{\frac{3\pi}{32G\rho_{\rm ta}}}\label{eq:tta}
\end{equation} is simply the freefall time of the shell. The above solution is then fully specified by two constants of integral, $\E$ and $t_0$, with $t_0=t(r=0)$ being the time when $r=0$. These two constants can be determined by specifying the initial conditions (IC), $r(t_\ri)$ and $\dot{r}(t_\ri)$ at a given initial time $t_{\ri}$. For example, it is often assumed that the perturbation is initially co-expanding with the background universe after recombination, with $\dot{r}(t_\ri)=H_\ri r(t_\ri)$. If $r(t_\ri)$ is also specified for the region of mass $M$, then $\E$ and $t_0$ are fully specified. 
However, as long as $t_0$ is small compared to the dynamical timescale for the system, for late time evolution we can safely adopt $t_0=0$. This corresponds to an initial condition under which the perturbed region started from a singularity, $r=0$, in the beginning of the universe. \citet{GunnGott72} provided more discussions on the different initial conditions and their influences. When studying the classical spherical collapse model we will mostly adopt the simple singular IC. Meanwhile, we will switch to the more realistic co-expansion IC when studying secondary infall models in section~\ref{sec:secondary}. 

\begin{figure}
    \includegraphics[width=\textwidth]{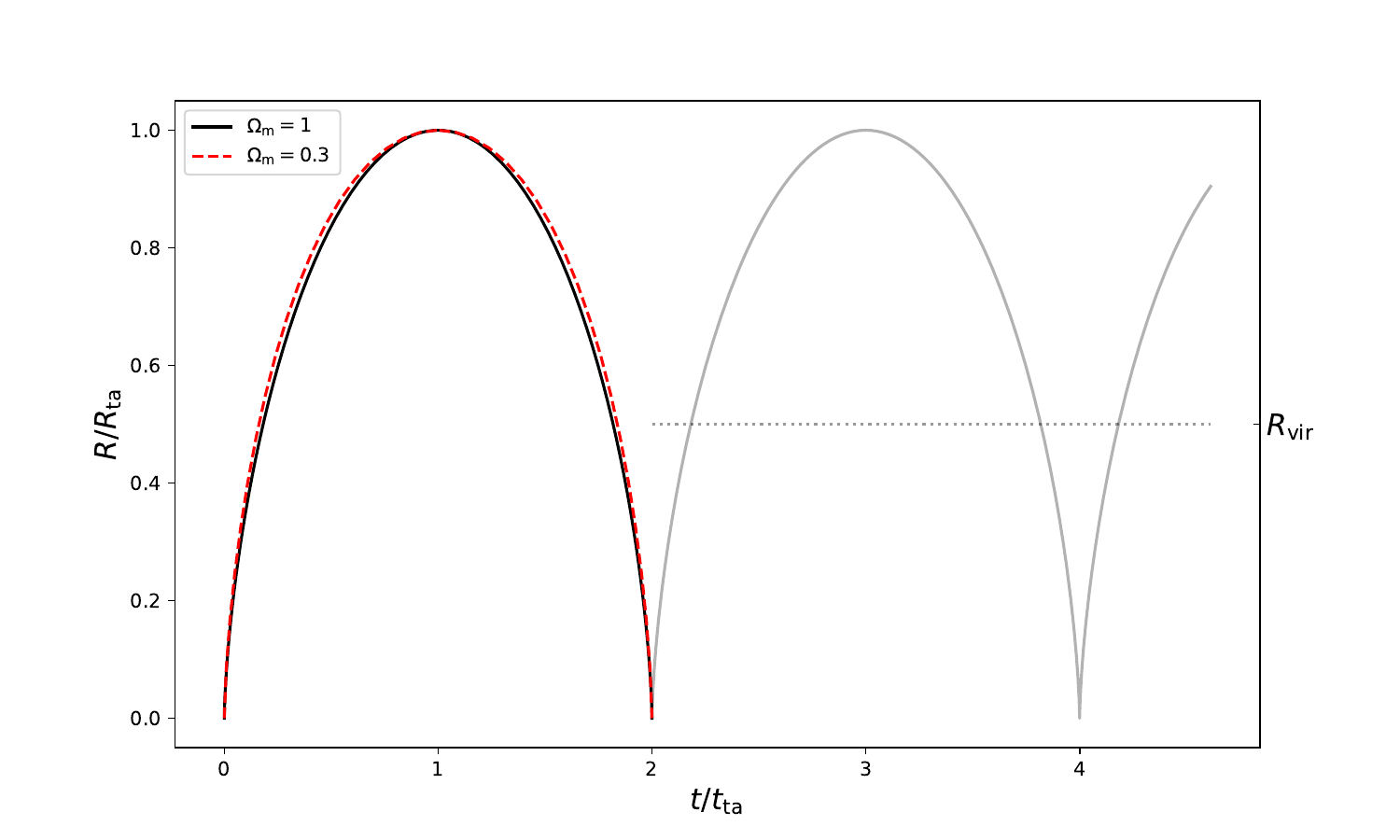}
    \caption{Monolithic spherical collapse solution. The shell oscillates periodically in the monolithic spherical collapse model, but will evolve towards virialization in reality, as indicated by the dotted curve. The black solid curve shows the solution in the EdS universe. For comparison, the red dashed curve shows the solution in a flat universe with parameters $\Omega_{\rm m}=0.3$ and $\Omega_{\Lambda}=0.7$ at the turnaround time.
    }\label{fig:sc_sol}
\end{figure}
The solution given in Equations~\eqref{eq:sc_sol_r} and \eqref{eq:sc_sol_t} describes a system that first expands to its maximum radius $r_{\rm ta}$ at $\phi=\pi$, and then falls back freely till $r=0$ at $\phi=2\pi$ (see Figure~\ref{fig:sc_sol}). The radius $r_{\rm ta}$ is thus named the \emph{turnaround} radius. The time at $\phi=2\pi$ is also called the collapse time, $t_{\rm c}=2t_{\rm ta}$. The spherical collapse model in the $\Lambda=0$ case is thus simply the freefall process of a bound system preceded by an inverse of this process, i.e., free expansion under its self-gravity. It is important to recognize that the turnaround time and radius are fully determined by the mass and energy of the shell, which in turn determines the complete orbit of the shell. Consequently, halos of the same mass can collapse at different times depending on their energy parameters.

Because the spherical collapse is a free-fall process, the turnaround density and time follow the free-fall scaling of $t_{\rm ta}\propto \rho_{\rm ta}^{-1/2}$. The same is true for the density evolution in an Einstein-de-Sitter (EdS) universe with $\Lambda=0$ and $k=0$. It can thus be understood that the turnaround density scales universally with the background density at turnaround time. More specifically, in a matter-only (with $\bar{\rho}\propto R^{-3}$) EdS universe, the growth of density can be found from Equation~\eqref{eq:Friedmann_vel} as
\begin{equation}
    t=\frac{1}{\sqrt{6\pi G\bar{\rho}}}.\label{eq:EdSdensity}
\end{equation}
Combined with Equation~\eqref{eq:tta}, the turnaround overdensity can be found as
\begin{equation}
\delta_{\rm ta}=\rho_{\rm ta}/\bar{\rho}(t_{\rm ta})-1=\frac{9\pi^2}{16}-1\approx 4.55\label{eq:delta_ta}
\end{equation}

\subsection{General solutions}
\subsubsection{Cosmological parameter form}
With $\rho_{\rm c}\equiv 3H^2/8\pi G$, $\Omega_{\rm m}\equiv\bar{\rho}/\rho_{\rm c}$, $\Omega_{\rm k}=-\frac{kc^2}{H^2 R^2}$ and $\Omega_\Lambda=\Lambda/3H^2$, the first Friedmann equation (Equation~\eqref{eq:Friedmann_vel}) can be rewritten as
\begin{equation}
    (\frac{\ud a}{\ud \tau})^2= a^{-1}+\omega a^{2} - \kappa,\\\label{eq:arate_reduce}
\end{equation}
where $\tau\equiv \sqrt{\Omega_{\rm m,i}}H_{\rm i} t$, $\omega\equiv \Omega_{\Lambda,\mathrm{i}}/\Omega_{\rm m,i}$ and $\kappa\equiv-\Omega_{k,\ri}/\Omega_{\rm m,i}$. The subscript $\mathrm{i}$ denotes quantities defined at a reference time, $t_{\rm i}$, and the dimensionless scale factor is defined as $a\equiv R/R_\ri$.


 Similarly, we can define a scale factor for the perturbed region as $a_{\rm p}\equiv r/r_{\rm L}$, where 
 \begin{equation}
 r_{\rm L}\equiv (1+\delta_{\rm i})^{1/3}r_{\rm i}=[M/(\frac{4\pi}{3}\bar{\rho}_{\rm i})]^{1/3}\label{eq:r_L}
 \end{equation}
 is the \emph{Lagrangian} radius of the perturbation.\footnote{As long as $\delta_\ri\ll1$, $r_{\rm L}\approx r_{\rm i}$ is a good approximation.} The spherical collapse energy equation (Equation~\eqref{eq:Newton_Vel}) then simplifies to
\begin{equation}
(\frac{\ud a_\rp}{\ud \tau})^2=a_\rp^{-1}+\omega a_\rp^{2} - \kappa_\rp. \label{eq:aprate_reduce}
\end{equation} where $\kappa_\rp\equiv-\Omega_{\rm kp,i}/\Omega_{m,\ri}$, and $\Omega_{\rm kp,i}\equiv \frac{2\E}{(H_\ri r_\mathrm{L})^2}$. Note the cosmological parameters $H_\ri$, $\Omega_{m,i}$ and $\Omega_{\Lambda,i}$ are all background ones, while the dependence on the energy $\E$ and mass $M$ (through $r_\mathrm{L}$) of the perturbation have both been absorbed into the $\Omega_{\rm kp,i}$ parameter. 

With the above definitions, the perturbed density evolves as
\begin{equation}
 \rho=\bar{\rho}_\ri r_\mathrm{L}^3/r^3=\bar{\rho}_\ri a_\rp^{-3}
\end{equation} and the perturbation overdensity is
\begin{equation}
 1+\delta\equiv\frac{\rho}{\bar{\rho}}=(1+\delta_\ri)(\frac{a}{a_\ri})^3(\frac{a_\rp}{a_{\rp,\ri}})^{-3}=(\frac{a}{a_\rp})^3.\label{eq:delta_p}
\end{equation} 
According to Equations~\eqref{eq:arate_reduce} and \eqref{eq:aprate_reduce}, $a_\rp$ expands slower than $a$ when $\kappa_\rp>\kappa$ (i.e., the perturbation is more bound than the background universe). In this case, $\delta$ will grow. On the other hand, $\kappa_\rp<\kappa$ corresponds to decaying perturbations that are less bound than the background. 


\subsubsection{Turnaround}
The cosmic time serves as a link between the perturbed region and the background universe. Turnaround happens at $\dot{a}_\rp=0$, or 
\begin{equation}
a_{\rm p,ta}^{-1}+\omega a_{\rm p,ta}^{2} =\kappa_\rp.\label{eq:ap_ta_kappa}
\end{equation} This is simply the energy conservation equation at turnaround time when the kinetic energy vanishes, leaving only the potential energies on the left-hand side. The parameter $\kappa_\rp$ can be recognized as the dimensionless binding energy. 
The turnaround time can be found as 
\begin{align}
\tau_{\rm ta}&=\int_0^{a_{\rm p,ta}} \frac{\ud \tau}{\ud a_\rp} \ud a_\rp\nonumber\\
&=\int_0^{a_{\rm p,ta}}\frac{\ud a_\rp}{\sqrt{a_\rp^{-1}+\omega a_\rp^{2} - \kappa_\rp}}.\label{eq:tau_p}
\end{align} Similarly,
\begin{align}
\tau_{\rm ta}&=\int_0^{a_{\rm ta}} \frac{\ud \tau}{\ud a} \ud a\nonumber\\
&=\int_0^{a_{\rm ta}}\frac{\ud a}{\sqrt{a^{-1}+\omega a^{2} - \kappa}}.
\end{align} Equating the two $\tau_{\rm ta}$, we can get the mapping between $a_{\rm ta}$ and $a_{\rm p,ta}$ or equivalently $\kappa_\rp$. The turnaround density is then given by Equation~\eqref{eq:delta_p} at turnaround time. We see that the turnaround time $\tau_{\rm ta}$ or equivalently $a_{\rm ta}$ is determined only by the energy or curvature parameter of the shell, $\kappa_\rp$, and vice versa, in line with the $\Lambda=0$ case. \citet{Eke96} provided an analytical solution for $a_{\rp,{\rm ta}}(\kappa_\rp)$ in a flat universe~\citep[see also][]{Shi16}.


Note the turnaround and subsequently virial quantities in general depends on the cosmology, $\Omega_\ri$, and time of the event, e.g., $a_{\rm ta}$. However, as we have the freedom to choose the reference time, we can choose it at the relevant time of interest, for example, at the turnaround time, so that $a_{\rm ta}=1$. Then the results such as turnaround density \emph{only depends on the cosmological parameters at turnaround}. The same argument applies to the virial quantities. 

The general solution to the full orbital evolution is also shown in Figure~\ref{fig:sc_sol} for a $\Omega_\Lambda=0.7$ flat universe. The shape of the solution deviates only weakly from the EdS solution, when normalized by the turnaround quantities.

\subsubsection{Maximum turnaround radius}
According to Equation~\eqref{eq:SC_Acc}, the acceleration of the mass shell originates from the competition between the gravity of the enclosed mass and a repulsive force due to $\Lambda$. While gravity decreases with radius, the repulsion increases over radius. The two cancels each other at a transition radius
\begin{equation}
 r_\Lambda\equiv\left(\frac{3GM}{\Lambda}\right)^{1/3}.\label{eq:rta_max}
\end{equation} Beyond $r_\Lambda$, dark energy takes over gravity, and a shell that expands past it will thus never collapse back. This radius is called the maximum turnaround radius.

\begin{figure}
\includegraphics[width=\textwidth]{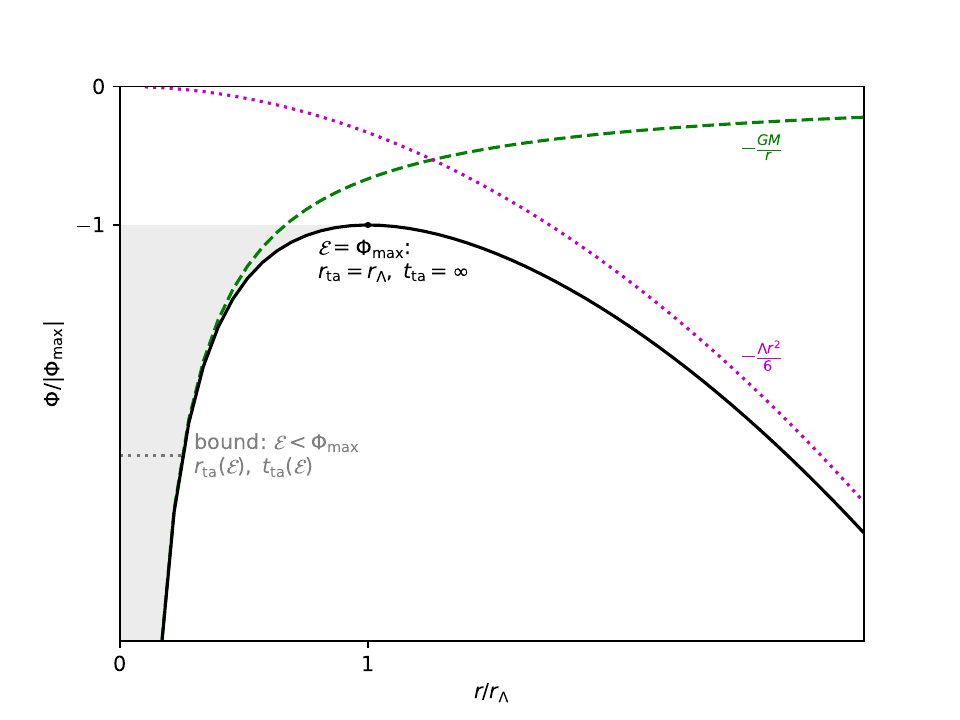}
\caption{The total potential for a spherical shell of mass $M$ in presence of the cosmological constant. The contributions from the Newtonian gravity and from the cosmological constant are shown by the green and magenta curves respectively. The total potential reaches its maximum at $r_\Lambda$, and the shell can ever turnaround (i.e., is bound) only when its total energy is below $\Phi_{\rm max}$.}\label{fig:max_turnaround}
\end{figure}

In energy argument, the potential energy of the system can be written as $\Phi=-\frac{\Lambda r_\Lambda^2}{6}[2/\tilde{r}+\tilde{r}^2]$, where $\tilde{r}\equiv r/r_\Lambda$. This potential energy reaches a maximum of $\Phi_{\rm max}=-\frac{\Lambda r_\Lambda^2}{2}$ at $r_\Lambda$. This means the condition for the system to be bound in presence of $\Lambda$ is no longer $\E<0$, but $\E<\Phi_{\rm max}$ (see Fig.~\ref{fig:max_turnaround}). 
The same analysis can be done with the dimensionless energy conservation equation, Equation~\eqref{eq:ap_ta_kappa}. In this representation, the maximum turnaround radius is expressed as $a_{p,\Lambda}=(2w)^{-1/3}$. The bound condition becomes 
\begin{equation}
    \kappa_\rp > \kappa_{\rm p,min}\equiv\frac{3}{2}(2w)^{1/3}.
\end{equation}

As the energy, turnaround radius and turnaround time have one-to-one mappings with each other for a given mass and cosmology, the maximum turnaround radius also corresponds to a maximum turnaround time, after which halos of the given mass can no longer form. For a critically bound shell with $\kappa_\rp= \kappa_{p,\rm min}$, one can expand the denominator in Equation~\eqref{eq:tau_p} around the critical turnaround radius $a_{\mathrm{p},\Lambda}$, to obtain $\ud \tau (a_{\rm p}) \propto \ud \ln (a_{\rm p,\Lambda}-a_{\rm p})$, so that integrating up to the maximum turnaround radius results in a trivial maximum turnaround time of $\tau_{\rm ta,max}=\infty$. This means that halo formation never completely halts, despite that later collapse becomes progressively slower as the turnaround radius approaches $r_\Lambda$. 

\subsection{Virialization}
The spherical collapse solution predicts the shell to collapse to $r=0$ and then bounces back to repeat the spherical collapse process. According to Equation~\eqref{eq:tau_p} (see also Equation~\eqref{eq:tta}), the turnaround time of a shell is determined by the energy parameter $\kappa_{\rm p}$. Under the co-expansion IC, it is easy to show that $\kappa_{\rm p}$ only depends on the initial overdensity (see Equation~\eqref{eq:energy-delta}). Thus an initially uniform spherical region will evolve monolithically. 

For non-uniform perturbed regions, however, different shells will turnaround and collapse asynchronously, leading to shell-crossing during their evolutions. Shell-crossing causes the enclosed mass of each shell to evolve over time, and the simple SC solution above no longer holds. For an initial density peak with a decreasing density profile, outer shells turnaround later, so that violation of the simple SC model only happens some time after turnaround. Shell-crossing results in energy exchange among the shells, eventually leading to virialization of the inner part of the perturbed region. Self-consistent solutions in presence of shell-crossing will be discussed in section~\ref{sec:secondary} below. For now, we will focus on the property of the final virialized region. 

For a uniform mass distribution in the perturbed region, the gravitational potential energy is
\begin{equation}
U_M(r)=-\frac{3}{5}\frac{GM^2}{r}.
\end{equation}
The potential energy due to $\Lambda$ is
\begin{equation}
    U_\Lambda(r)=-\frac{\Lambda}{10}Mr^2.
\end{equation} Using the virial theorem $T=\dfrac{n}{2}U$ for $U\propto r^n$~\citep{Landau}, the final kinetic energy at virial equilibrium can be found as~\citep{Lahav91}
\begin{equation}
    T_v=-\frac{1}{2}U_M(r_v)+U_\Lambda(r_v).\label{eq:virial}
\end{equation} Using energy conservation to relate the total final energy to the potential energy at turnaround, an approximate solution to the final radius can be found as~\citep{Lahav91}
\begin{equation}
    \frac{r_v}{r_{\rm ta}}\approx \frac{1-\eta/2}{2-\eta/2},\label{eq:rv}
\end{equation} where $\eta\equiv \dfrac{\Lambda}{4\pi G\rho_{\rm ta}}$. This solution is applicable for $r_v/r_{\rm ta}=1/2+\epsilon$ with a small $\epsilon$. When $\eta=0$, the exact relation $r_v=\dfrac{1}{2}r_{\rm ta}$ is recovered.


Once $r_v$ is known, the virial density can be obtained from $\rho_{\rm vir}=\rho_{\rm ta}r^3_{\rm ta}/r^3_v$ while $\rho_{\rm ta}$ can be obtained from $a_{p,{\rm ta}}$. Assuming virialization happens at twice the turnaround time, $t_{\rm c}=2t_{\rm ta}$, the background density at virialization can also be obtained. For $\Lambda=0$, we have $\rho_{\rm vir}=8\rho_{\rm ta}$. From turnaround to collapse time the background density has evolved as $\bar{\rho}(t_c)=\bar{\rho}(t_{\rm ta})/4$ in an EdS universe (see Equation~\eqref{eq:EdSdensity}). Now the virial density contrast can be found as
\begin{equation}
    \Delta_{\rm vir}\equiv \frac{\rho_{\rm vir}}{\bar{\rho}(t_c)}=32\rho_{\rm ta}/\bar{\rho}(t_{\rm ta})=18\pi^2\approx 178, \label{eq:virial_delta_EdS}
\end{equation} where we have made use of Equation~\eqref{eq:delta_ta} for the turnaround density. For more general cosmologies, \citet{BryanNorman98} has fitted the derived virial density contrast as a function of the cosmological parameters at virialization time following the calculations in~\citet{LC93} and \citet{Eke96}, as
\begin{equation}
\Delta_{\rm c}=\begin{cases}
18\pi^2+82x-39x^2 &\quad  (\Omega_k=0),\\
18\pi^2+60x-32x^2 &\quad  (\Omega_\Lambda = 0),
\end{cases}\label{eq:BryanNorman}
\end{equation} where $x=\Omega_{\rm m}(z)-1$. This equation is accurate to 1\% in the range $\Omega_{\rm m}(z)=0.1-1$. 

Note such a density is derived with quite a few assumptions which may not be very well tested. For example, it assumes the virialization happens at a time of $2t_{\rm ta}$, while early simulation results from \citet{Peebles70} indicate a virialized halo at $3t_{\rm ta}$.
The potential energies in the virial theorem have been derived assuming a uniform mass distribution,

while realistic halos follow the Navarro-Frenk-White~\citep[NFW;][]{nfw96,nfw97} type of density profile. In addition, some external mass shells could also have entered inside the virial radius, violating mass conservation in the above derivations. Because the virialized part is embedded in a non-virialized environment with mass exchange at the boundary, the virial theorem (Equation~\eqref{eq:virial}) should also contain a boundary term. 
Due to these limitations, the virial density for the EdS universe is commonly used as a reference value, and simplified definitions such as $\rho_{\rm 200m}=200\bar{\rho}$ and $\rho_{\rm 200c}=200\rho_{\rm crit}$ are often adopted. Nevertheless, a characteristic density contrast of the virialized halo arises naturally from the shared dynamical timescale of  halo collapse and background expansion, irrespective of the detailed evolution paths. Thus it remains generally meaningful to characterize the collapse region with the virial density contrast. Some simulation tests to the model are presented in Section~\ref{sec:simu_test}.

\subsection{Linear evolution and extrapolated properties}\label{sec:linear_collapse}
It is often useful to consider the evolution of the perturbation in the linear theory, so that halo statistics could be derived from the linearly evolved density field. For simplicity, we will start from the EdS universe. The non-linear evolution of the overdensity of the perturbed region is given by
\begin{align}
	1+\delta&=(1+\delta_{\rm ta})(\frac{r_{\rm ta}}{r})^3(\frac{a}{a_{\rm ta}})^3\\
	& =\frac{9\pi^2}{16}(\frac{r_{\rm ta}}{r})^3(\frac{a}{a_{\rm ta}})^3\\
 & =\frac{9}{2}\frac{(\phi-\sin\phi)^2}{(1-\cos\phi)^3},
\end{align} 
where the last equality is obtained by substituting the solutions for $r$ and $t$ along with $a\propto t^{2/3}$ in an EdS universe.

For a small $\phi$, we can expand the solution to leading order in $\phi$, to obtain the linear growth of the density field,
\begin{equation}
	\delta=\frac{3}{20}\phi^2.
\end{equation} Combined with the leading order expansion of time,
\begin{equation}
	t=t_{\rm ta}\frac{\phi^3}{6\pi},
\end{equation} the time evolution of $\delta$ is
\begin{align}
	\delta&=\frac{3}{20}(\frac{6\pi t}{t_{\rm ta}})^{2/3}\\
	&\propto t^{2/3}\propto a.
\end{align} This recovers the linear growth rate in the EdS universe. 

Note the above equation applies only to a small $\phi$ and hence small $t$ and $\delta$. Nevertheless, if we evolve the density field according to the linear growth rate, then for any perturbation the evolved density will always reach a constant at its turnaround time, 
\begin{equation}
	\delta_{\rm ta}^{\rm lin}\equiv \frac{3}{20}(6\pi)^{2/3}\approx 1.06.
\end{equation} Similarly, extrapolating to the collapse time, $t=2t_{\rm ta}$, we have the linear theory prediction of the virial density contrast,
\begin{equation}
	\delta_{\rm sc}^{\rm lin}=\delta_{\rm ta}^{\rm lin}2^{2/3} \approx 1.686.
\end{equation} These results are particularly useful for identifying collapsed regions in a linearly evolved density field, in particular in the extended Press-Schechter (EPS) approach to halo statistics (see section~\ref{sec:eps} below).


For general cosmologies, expanding the nonlinear evolution of Equation~\eqref{eq:delta_p} to linear order in the scalefactor and comparing against the linear theory prediction, 
one can connect the perturbation parameter $\kappa_\rp$ to the linear perturbation at a reference time, e.g., present day~\citep{Eke96, PF05}. 

The linearly extrapolated overdensity at turnaround and virialization can be then obtained according to the growth rate. Alternatively, one can work with the differential equations on the evolution of overdensity directly~\citep{Pace10,Mead17}. Taking the derivatives of Equation~\eqref{eq:delta_p} and making use of Equation~\eqref{eq:SC_Acc} and \eqref{eq:Friedmann_Acc}, the nonlinear evolution of $\delta$ is given by
\begin{equation}
    \ddot{\delta}=-2H\dot{\delta}+\frac{4}{3}\frac{\dot{\delta}^2}{1+\delta}+4\pi G\bar{\rho}(\delta+\delta^2).\label{eq:delta_nonlin}
\end{equation} To linear order, the above equation simplifies to
\begin{equation}
    \ddot{\delta}=-2H\dot{\delta}+4\pi G\bar{\rho}\delta.\label{eq:delta_lin}
\end{equation} Evolving Equation~\eqref{eq:delta_nonlin} with different initial conditions till $\delta=\infty$,

the corresponding solution to Equation~\eqref{eq:delta_lin} at the same $t$ then gives the linear overdensity at collapse. For general cosmologies, these values are cosmology and redshift dependent. Fitting formulas for $\delta_c$ (and $\Delta_{\rm vir}$) have been provided in ~\citep{NS97, Mead17}. For $\Omega_{\rm m}+\Omega_{\Lambda}=1$, the linear collapse density can be fitted to within 0.1\% for $\Omega_v>0.01$ as \citep{NS97}
\begin{equation}
    \delta_{\rm c} \simeq \frac{3}{20}(12\pi)^{2/3}(1+0.012299\log \Omega_{\rm v}),
\end{equation} where $\Omega_{\rm v}$ is the matter density parameter at the virialization time.

\subsection{Simulation tests}\label{sec:simu_test}
The analytical model above provides the theoretical foundation for defining the boundary of a halo in simulations and observations, and results in an industry of studies on halo statistics in numerical simulations~\citep[see][for reviews]{CooraySheth02,Zavala19Rev,AnguloHahn22} often accompanied by theoretical models exploiting the statistics of density peaks in the linear density field~\citep{PS,BBKS,BCEK,LC93,PeakPatch,CUSP}.

\begin{figure}
    \includegraphics[width=0.49\textwidth]{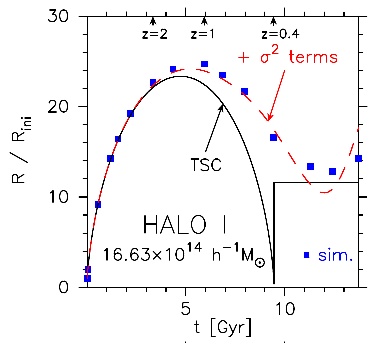}
    \includegraphics[width=0.49\textwidth]{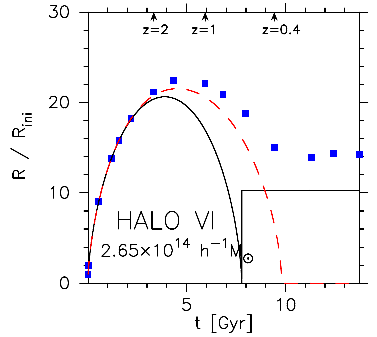}
    \caption{Evolution of shell size compared with model predictions. The data points show the evolution of the size of a spherical region of a fixed enclosed mass in simulation. The solid and dashed curves are the predicted size evolution from the top-hat spherical collapse model and the non-stationary Jeans equation, respectively. The left and right panels are two different example halos. Figure reproduced from \citet{Suto16}.}\label{fig:SCtest}
\end{figure}

To test the validity of the virial radius definition, \citet{Cuesta08} computed the \emph{static radius} of each halo in cosmological simulations, defined as the radius within which the net radial velocity profile is approximately zero, a necessary condition for virialization. They found that the static radius is generally larger than the virial radius, especially for low mass halos. At $z=0$, the static mass can be as large as two times the virial mass for $M_{\rm vir}\sim 10^{12}M_\odot$ halos, while the two become close in cluster size halos. \citet{Zemp14} studied the evolution of the halo mass profile and reached similar conclusions, finding that the mass shells beyond the virial radius are not showing much evolution in the last few Gyr except for the most massive objects. In low mass halos, the evolution of the virial radius within the static region is completely driven by the evolution of the background density, rather than reflecting any ongoing structural evolution, and is thus called pseudo-evolution~\citep{PseudoEvo}. \citet{Cuesta08} attributes the mismatch between the static radius and the SC expectation to the effect of velocity dispersion, generated by the tidal influence of neighboring halos. By tracking the evolution of spherical shells in simulated halos, \citet{Suto16} directly verified that the spherical collapse model generally works till turnaround (see Figure~\ref{fig:SCtest}), after which the evolution is better modeled with the non-stationary Jeans equation that accounts for the velocity dispersion of particles in addition to gravity,\footnote{The $\Lambda$ term is not included in the original work of \citet{Suto16}}
\begin{equation}
\ddot{r}=-\frac{GM}{r^2}+\frac{\Lambda}{3}r-\frac{1}{\rho}\frac{\partial(\rho \sigma^2_r)}{\partial r}-\frac{2\sigma^2_r-\sigma^2_t}{r},\label{eq:Jeans}
\end{equation} where $\sigma^2_r=<v^2_r>$ and $\sigma_t=<v^2_t>$ are the radial and tangential velocity moments.

Even though the virial radius is located inside the static region, it remains to be tested whether virial equilibrium is satisfied inside it. For an isotropic velocity distribution with $\sigma_{\rm tot}^2=3\sigma_r^2$, one can multiply the stationary ($\ddot{r}=0)$) version of the Jeans equation (Equation~\eqref{eq:Jeans}) by $4\pi r^3\ud r$ and integrate out to the virial radius, to find
\begin{equation}
   E_{\rm s}-2T=U_M(r_v)-2U_\Lambda(r_v). \label{eq:virial_Es}
\end{equation} Here $T=\frac{1}{2}\int_0^{r_v} \sigma^2_{\rm tot}\ud m$, $U_M=-\int_0^{r_v} \frac{GM(r)\ud m}{r}$, and $U_\Lambda=\int_0^{r_v} \frac{\Lambda r^2}{6}\ud m$ are the kinetic energy and potential energies due to gravity and $\Lambda$, respectively. $U_\Lambda$ is usually very small compared with $U_{M}$, for example, less than 1\% at $z=0$ for the concordance cosmology, and can be safely ignored~\citep{Suto16}. The surface pressure term, 
\begin{equation}
    E_{\rm s}\equiv \frac{4\pi}{3}r_v^3\rho(r_v)\sigma^2_{\rm tot}(r_v)
\end{equation} arises due to random motions at the boundary of an embedded system~\citep{Chandrasekhar1961,Carlberg96}. For isolated systems, Equation~\eqref{eq:virial_Es} reduces to Equation~\eqref{eq:virial}. \citet{Shaw2006} showed that the virial theorem is approximately satisfied by simulated cluster halos out to the virial radius, with a virial ratio $\eta\equiv (2T-E_s)/|U_M|$ close to 1 and a scatter in $\eta$ of $\sim 0.1$. The virial ratio increases weakly with halo mass~\citep{Power12} and redshift~\citep{Davis11}, with massive and high redshift halos relatively less relaxed on average. Results from hydrodynamical simulations show that baryons tend to relax slightly earlier than the DM component~\citep{Poole06}, leading to $\sim 10\%$ decrease in the virial ratio of $z=0$ clusters when baryons are included~\citep{Cui17}. 

The virialization state of halos has also been examined in the context of dynamical mass measurement. Using a minimal assumption dynamical modeling method~\citep{oPDF} that essentially only assumes the halo to be in a steady-state, \citet{oPDF2} and \citet{Wang17} showed that the virial mass of galactic halos in simulations can be inferred with a systematic uncertainty of $\sim 0.1$ dex, while the uncertainty increases to $\sim 0.2$ dex in cluster halos~\citep{Li21,Li22}. Satellite galaxies as dynamical tracers virialize to a similar level as DM~\citep{Han20}. The gas component can be more virialized~\citep{Li21} than the DM, while halo stars are the least virialized~\citep{Han20}. These uncertainties arise due to the approximate virialization state of the halo, and thus is shared by dynamical models making the steady-state or virial assumption~\citep[see e.g.,][for tests of other methods and tracers]{Wang15,Wang18,Li19,Old15}.

\subsection{Connection to halo statistics}\label{sec:eps}
The abundance of halos as a function of halo mass, i.e., the halo mass function, can be generally understood from statistics of the Gaussian density field describing the initial condition of structure formation. As introduced in Section~\ref{sec:linear_collapse}, collapsed halos at a given epoch can be equivalently identified as peaks containing the given mass that reach the critical overdensity, $\deltac$, in the linearly evolved density field. The statistics of peaks containing a certain mass, $M$, are specified by the Gaussian density field smoothed on the corresponding perturbation scale, $r_L(M)$,  given in Equation~\eqref{eq:r_L}. The abundance of peaks reaching $\deltac$ is then only determined by the peak height,
\begin{equation}
    \nu(M)\equiv \frac{\deltac}{\sigma(M)},
\end{equation} with $\sigma(M)$ being the standard deviation of the Gaussian field smoothed on the scale $r_L(M)$ and linearly evolved to the collapse time. Consequently, the fractional contribution of halos in a given mass interval, $[M, M+\ud M]$, to the matter density can be expressed through a dimensionless function, $f(\nu)$, usually defined as
\begin{equation}
f(\nu) \ud \nu \equiv \frac{M\ud n(M)}{\bar{\rho}}.
    \label{eq:num_dens}
\end{equation} 
The function $f(\nu)$ is predicted to be a Gaussian function in the original excursion set models~\citep{PressSchechter, BCEK, LC93}, independent of cosmology and redshift. 

\citet{sheth1999large} modified $f(\nu)$ to better match simulation results as
\begin{equation}
    f_{\rm ST}(\nu)=A\sqrt{\frac{a}{2\pi}} [1+(a\nu^2)^{-p}]e^{-\frac{a\nu^2}{2}},
    \label{eq:ST}
\end{equation} with $A=0.322$, $p=0.3$ and $a^2=0.707$, while the original excursion set prediction corresponds to $A=1/2$, $p=0$ and $a=1$. The modification introduced in the Sheth-Tormen mass function can be interpreted as arising from ellipsoidal collapse, which results in a moving barrier, $\deltac(M)$, instead of the constant barrier predicted by the spherical collapse model~\citep{sheth2001ellipsoidal}. 

The largescale clustering of halos, or equivalently, the linear bias, defined as the ratio between the overdensity of halos to that of mass, $b=\delta_h/\delta_m$, can also be derived from the statistical properties of the initial Gaussian random field, by investigating the environmental dependence of the halo-forming peaks~\citep{BBKS, CK89, MW96}. The general dependence is that more massive halos are more highly biased. 

These halo statistics are generally developed based on the virial-halo definition. However, the general formalisms are found to also work for other halo definitions developed recently, which we will discuss in more detail in section~\ref{sec:population_stat}.

\section{Dynamical boundaries around a growing halo}
\label{sec:secondary}

The spherical collapse model not only applies to the collapse of the perturbed region, but also applies to the ``secondary" infall of the surrounding mass onto an already collapsed structure, which was in fact the main focus in the majority of early works. The collapse time of each mass shell and the corresponding enclosed mass then leads to a model for the mass accretion rate~\citep{GunnGott72}. The density profile of the extended halo could also be derived by considering how each shell contributes to the final mass distribution~\citep{Gott75,Gunn77}. To achieve this, it becomes necessary to consider interactions among different mass shells when evolving their orbits. According to Equation~\eqref{eq:aprate_reduce}, the collapse of different shells is asynchronous unless they have identical energy parameters, $\kappa_\rp$. Thus different shells will generally exchange positions after turnaround, modifying the enclosed mass of each shell over time and thus invalidating the monolithic spherical collapse calculation.

To account for shell-crossing, one can in principle appeal to numerical simulation of multiple shells~\citep[e.g.,][]{Gott75}. Such calculations can be greatly simplified if the system is self-similar, such that different shells all follow the same orbit once properly scaled. In this case, the density and thus potential of the perturbed region is also self-similar, and can be solved self-consistently with the orbit of the shell. \citet{Gunn77} first exploited self-similarity to investigate the relaxed halo profile approximately. \citet{FG84} first made a complete calculation of self-similar secondary collapse in the EdS universe. \citet{Bertschinger85} obtained self-similar solutions for both dark matter and gas infall, which can also be applied to the accretion around a black hole. \citet{Shi16, Shi23} extended the calculation to a more general case in which the density profile is not self-similar but follows the NFW profile. These models are particularly useful for understanding the non-equilibrium region around a halo characterized by the dynamical boundaries such as the splashback and depletion radii.


\subsection{Co-expansion IC}\label{sec:coexp_ic}
In an unperturbed EdS universe, the Hubble flow and the potential energy are connected as (Equation~\eqref{eq:Friedmann_vel})
\begin{equation}
	\frac{1}{2}H^2r^2=\frac{G\bar{M}}{r}.
\end{equation}

The co-expansion IC for spherical collapse assumes all perturbed regions expand at the same velocity as the background universe at an initial time, $t_\ri$, taken as the matter-radiation decoupling epoch~\citep{GunnGott72}, with
\begin{align}
r(t_\ri)&=r_\ri,\\
\dot{r}(t_\ri)&=H_\ri r_\ri.
\end{align}
Now the energy of the perturbed region can be written as 
\begin{align}
\mathcal{E}&=\frac{1}{2}(H_\ri r_\ri)^2-\frac{GM}{r_\ri}\\
&=-\frac{G\bar{M}_\ri\delta_\ri}{r_\ri}\label{eq:energy-delta}
\end{align} 
Noting that $M=\bar{M}_\ri(1+\delta_\ri)$, the turnaround quantities (Equations~\eqref{eq:rta} and \eqref{eq:tta}) can be related to the initial density as\footnote{Here $t_{\rm max}$ is used in place of $t_{\rm ta}$, with $t_{\rm ta}=t_0+t_{\rm max}\approx t_{\rm max}$ for halo collapsing late (see discussions in section~\ref{sec:sc_sol_eds}). }
\begin{align}
\frac{r_{\rm ta}}{r_\ri}&=\frac{1+\delta_\ri}{\delta_\ri}\label{eq:rta_delta_i}\\
t_{\rm ta}&=\frac{1}{H_\ri}\frac{\pi}{2}\frac{1+\delta_\ri}{\delta_\ri^{3/2}}\\
&=t_\ri\frac{3\pi}{4}\frac{1+\delta_\ri}{\delta_\ri^{3/2}},\label{eq:tta_delta_i}
\end{align} 
where we have used the relation $t=\frac{2}{3}\frac{1}{H}$ for the EdS universe. Note Equations~\eqref{eq:rta_delta_i} to \eqref{eq:tta_delta_i} are exact, not limited to a small $\delta_\ri$. 

The energy parameter $\kappa_{\rm p}$ can also be expressed in terms of the initial condition as 
\begin{equation}
    \kappa_{\rm p}=\frac{\delta_\ri}{(1+\delta_\ri)^{2/3}},
\end{equation} which is valid in a flat universe even in the presence of $\Lambda$. Note this $\kappa_\rp$ is defined at the co-expansion time. 

\subsection{Self-similar spherical collapse model}\label{sec:selfsimilar-sol}
\subsubsection{Perturbation profile and turnaround history}
Consider a power-law initial perturbation profile of the form
\begin{equation}
\delta_\ri\equiv \frac{\delta M_\ri}{M_\ri}=\frac{\delta \rho}{\rho}=(\frac{M_\ri}{M_0})^{-\epsilon}.\label{eq:FG_delta_ini}
\end{equation} We use $M_\ri$ to denote the initial mass, which also serves as a Lagrangian coordinate for the mass shell. The parameters $M_0$ and $\epsilon$ specify the scale and slope of the powerlaw. Together with the co-expansion IC, the energy profile of the perturbation is then fully specified. We restrict our discussions to $\epsilon>0$, corresponding to perturbations around density peaks. \citet{Bertschinger85} studied the collapse of background material around a fixed mass excess $\delta M_\ri=\mathrm{const} \ll M_\ri$, corresponding to the $\epsilon=1$ case here. 

Before shell-crossing, the mass $M_\ri$ within each shell is conserved. According to the standard spherical collapse (Equations~\eqref{eq:rta_delta_i} and \eqref{eq:tta_delta_i}), turnaround for shell $M_\ri$ can be found up to leading order in $\delta_\ri$ ($\delta_\ri \ll 1$) as\footnote{The definition of $\delta$ in Equation~\eqref{eq:FG_delta_ini} and in Equation~\eqref{eq:rta_delta_i} are slightly different, but up to leading order they produce the same equations for the turnaround quantities.}
\begin{equation}
r_{\rm ta}=r_\ri \frac{1}{\delta_\ri}
\end{equation}
and
\begin{equation}
t_{\rm ta}=t_\ri (\frac{C_t}{\delta_\ri})^{3/2},\label{eq:tta_Mi}
\end{equation}
where $C_t=(\frac{3\pi}{4})^{2/3}$, and $r_\ri(M_\ri)$ is the radius of the shell at the initial time, $t_\ri$. As $\epsilon>0$, inner shells turnaround earlier.

Replacing $r_\ri$ with the initial mass and density, we get\footnote{Note \citet{FG84} defined $M$ as the mass per solid angle, which correspond to $M/4\pi$ in our equations.}
\begin{align}
r_{\rm ta}&=\left[\frac{3M_\ri}{4\pi\rho_\ri}\right]^{1/3}(\frac{M_\ri}{M_0})^\epsilon \nonumber\\
&\simeq \left[\frac{3M_\ri}{4\pi\bar{\rho}(t_\ri)}\right]^{1/3}(\frac{M_\ri}{M_0})^\epsilon,\label{eq:rta_Mi}
\end{align} where the second equality is up to leading order in $\delta_\ri$. 

Combining Equations~\eqref{eq:FG_delta_ini}, \eqref{eq:tta_Mi} and \eqref{eq:rta_Mi}, the turnaround radius of the perturbed region evolves over time as 
\begin{equation}
R_{\rm ta}(t)=\frac{1}{C_t^{1+1/3\epsilon}}[\frac{3M_0}{4\pi\bar{\rho}(t_\ri)}]^{1/3}(t/t_\ri)^{2/3+2/9\epsilon},
\end{equation} and the corresponding turnaround mass growth history of the halo is
\begin{equation}
\frac{M_{\rm ta}(t)}{M_0}=\frac{1}{C_t^{1/\epsilon}}(\frac{t}{t_\ri})^{2/(3\epsilon)}.\label{eq:Mta}
\end{equation} This growth history is due to the assumed initial profile, which should at least be applicable over a short timescale during which the growth history can be approximated as a power-law. From now on, we will use $R_{\rm ta}$ to denote the turnaround radius of the halo at the time of interest, $R_{\rm ta}(t)$.

\subsubsection{Equation of motion}
Assuming self-similarity, the mass profile can be rescaled as
\begin{equation}
M(r,t)=M_{\rm ta}(t) \mathcal{M}(r/R_{\rm ta}).\label{eq:Mprof}
\end{equation} The dimensionless mass profile can be found by integrating over the mass of all shells within $r$, 
\begin{equation}
\mathcal{M}(r/R_{\rm ta})=\int_0^\infty \frac{dM_\ri'}{M_{\rm ta}} H[r-r'(t; M_\ri')],
\end{equation} where $H[u]=0$ for $u<0$ and $H[u]=1$ for $u\geq 0$ is the Heaviside step function, and $r'(t; M_\ri')$ describes the orbital evolution of shell $M_\ri'$. 

For each mass shell $M_\ri$, we can further define its dimensionless radius and time as
\begin{equation}
\lambda(M_\ri)=\frac{r}{r_{\rm ta}}, \tau=\frac{t}{t_{\rm ta}(M_\ri)}.
\end{equation} 
The halo turnaround radius can also be normalized as
\begin{equation}
\Lambda(\tau)\equiv\frac{R_{\rm ta}(t)}{r_{\rm ta}}= \tau^{2/3+2/(9\epsilon)}.
\end{equation} 
Combined with the dimensionless mass profile, we will see below that the equation of motion for shell $M_\ri$ becomes independent on $M_\ri$ when expressed in these dimensionless coordinates. This means the entire system is self-similar.

At a given time $t$, $M_\ri$ can be expressed as a function of $\tau$ due to the $t_{\rm ta}(M_\ri)$ dependence (Equation~\eqref{eq:tta_Mi} and \eqref{eq:FG_delta_ini}),
 \begin{align}
 \frac{M_\ri}{M_0}&=\frac{1}{C_t^{1/\epsilon}}(\frac{t_{\rm ta}}{t_\ri})^{2/(3\epsilon)}\nonumber\\
 &=\frac{1}{C_t^{1/\epsilon}}(\frac{t}{t_\ri\tau})^{2/(3\epsilon)}\label{eq:Mi_t}
 \end{align} Here $\tau$ is the relative time with respect to $t_{\rm ta}$, or the orbital phase of the shell. If the shell is right at its first turnaround phase, then $\tau=1$, and we recover Equation~\eqref{eq:Mta}. Combining the two, Equation~\eqref{eq:Mi_t} can be simplified as
\begin{equation}
\frac{M_\ri(\tau;t)}{M_{\rm ta}(t)}=\tau^{-2/(3\epsilon)}.\label{eq:Mi_Mta}
\end{equation} Equation~\eqref{eq:Mi_Mta} specifies how the initial mass is redistributed along the orbit.
The mass profile can then be rewritten through the change of variable from $M_\ri$ to $\tau$,
\begin{equation}
\mathcal{M}(\frac{\lambda}{\Lambda})=\frac{2}{3\epsilon}\int_1^\infty \frac{\ud \tau}{\tau^{1+2/(3\epsilon)}}H[\frac{\lambda}{\Lambda}-\frac{\lambda'(\tau)}{\Lambda'(\tau)}] \label{eq:SS_mass_prof}
\end{equation}
Considering the scaled orbit, $r/R_{\rm ta}=\lambda(\tau)/\Lambda(\tau)$, is a quasi-periodic oscillation (see Fig.~\ref{fig:ss_solution} for example), the mass integral can be more explicitly written as the summation of the segments inside $r(\tau)$, 
\begin{align}
\mathcal{M}(x)
&=\sum_{\frac{\lambda'(\tau)}{\Lambda'(\tau)}<x}\int \ud \tau^{-2/(3\epsilon)}\\
&=-\sum_{\frac{\lambda(\tau)}{\Lambda(\tau)}=x}\frac{1}{\tau^{2/(3\epsilon)}\mathrm{sign}(\dot{x})}\label{eq:SS_mass_prof_sum}
\end{align} where $x=\dfrac{\lambda}{\Lambda}=\dfrac{r}{R_{\rm ta}}$, and $\mathrm{sign}(\dot{x})=\mathrm{sign}(\frac{\dot{\lambda}}{\lambda}-\frac{\dot{\Lambda}}{\Lambda})$ gives the sign of the radial velocity. 

Now the equation of motion can be written as
\begin{equation}
\frac{\ud^2 \lambda}{\ud \tau^2}=-\frac{\pi^2}{8}\frac{\tau^{2/(3\epsilon)}}{\lambda^2}\mathcal{M}(\frac{\lambda}{\Lambda}).\label{eq:SS_EoM}
\end{equation} This is the acceleration equation in a growing potential, with $M_{\rm ta}\propto \tau^{2/3\epsilon}$. As it does not depend on $r_{\rm ta}$ or $t_{\rm ta}$, the orbits of all shells are now unified. The scaled time $\tau$ can be understood as specifying the phase on the orbit. 

In the presence of angular momentum $J$, we can define $J_{\rm ta}\equiv r_{\rm ta}^4/t_{\rm ta}^2$ for each shell~\citep{Bertschinger85} and $j\equiv J/J_{\rm ta}$, so that the equation of motion can be generalized to
\begin{equation}
\frac{\ud^2 \lambda}{\ud \tau^2}=-\frac{\pi^2}{8}\frac{\tau^{2/(3\epsilon)}}{\lambda^2}\mathcal{M}(\frac{\lambda}{\Lambda})+\frac{j^2}{\lambda^3}.\label{eq:SS_EoM_j}
\end{equation}
In the following, we will focus on the $j=0$ case as the standard self-similar collapse model. On the other hand,$j\ll1$ provides a good approximation which is much easier to solve numerically, as it avoids the force divergence at $\lambda=0$.

\subsubsection{Solution}\label{sec:ss_solution}

The boundary condition is specified at the turnaround time as
\begin{equation}
\lambda(1)=1,\; \frac{\ud \lambda(1)}{\ud \tau}=0.
\end{equation} Note the solution for $\tau<1$ is given by Equations~\eqref{eq:sc_sol_r} and \eqref{eq:sc_sol_t}.

Because the position $\lambda$ depends only on $\tau$ and $\lambda(\tau)$ is the same for any shell, the solution $\lambda(\tau)$ can be interpreted in two ways. For a given shell $M_\ri$, it represents the orbit of the shell, for which $r_{\rm ta}$ and $t_{\rm ta}$ are fixed while $t$ is varying. At a fixed $t$, $\lambda(\tau)$ describes the distribution of shells.
Explicitly, normalized by the current turnaround quantities,
\begin{align}
\frac{r(\tau)}{R_{\rm ta}}&=\frac{\lambda(\tau)}{\Lambda(\tau)}\\
\frac{v_r(\tau)}{V_{\rm ta}}&=\frac{\ud \lambda/\ud \tau}{\Lambda/\tau}
\end{align} where $v_r=\ud r/\ud t$ and $V_{\rm ta}\equiv R_{\rm ta}/t$.

Equation~\eqref{eq:SS_EoM} can be solved numerically once a mass profile $\mathcal{M}(x)$ is known, and in turn the solution can be used to build up the mass profile according to Equation~\eqref{eq:SS_mass_prof}. This process can be done iteratively till the mass profile converges.
In Appendix~\ref{app:practical_eqs}, we provide some practical equations for the evaluation of various profiles of the model. The \spheric package implementing the solutions is presented in Appendix~\ref{app:spheric}.


\begin{figure}
    \includegraphics[width=0.49\textwidth]{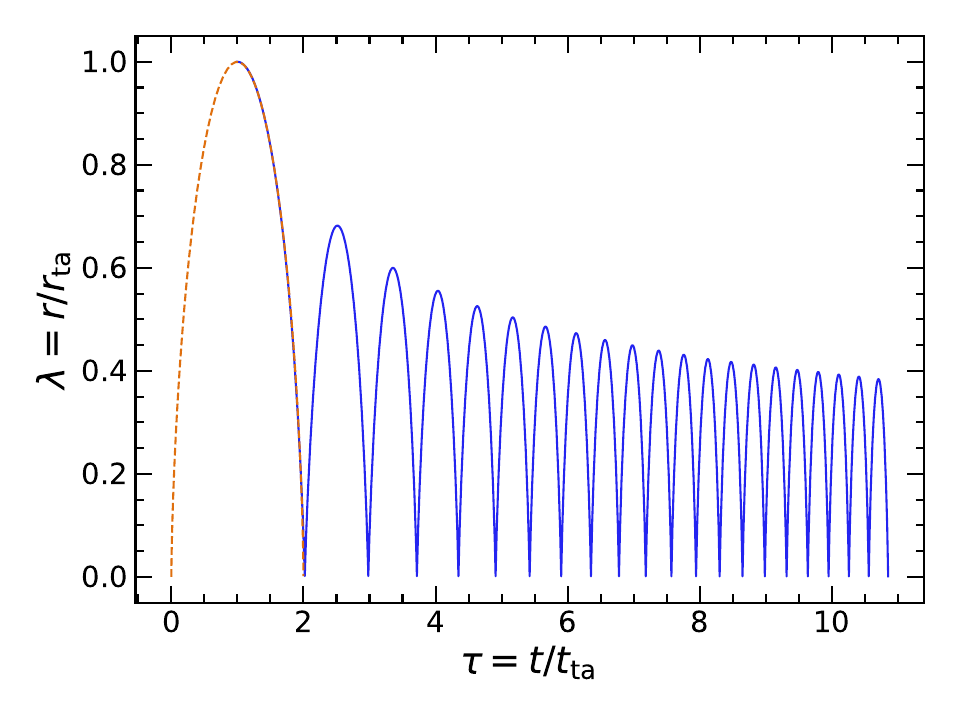}%
    \includegraphics[width=0.49\textwidth]{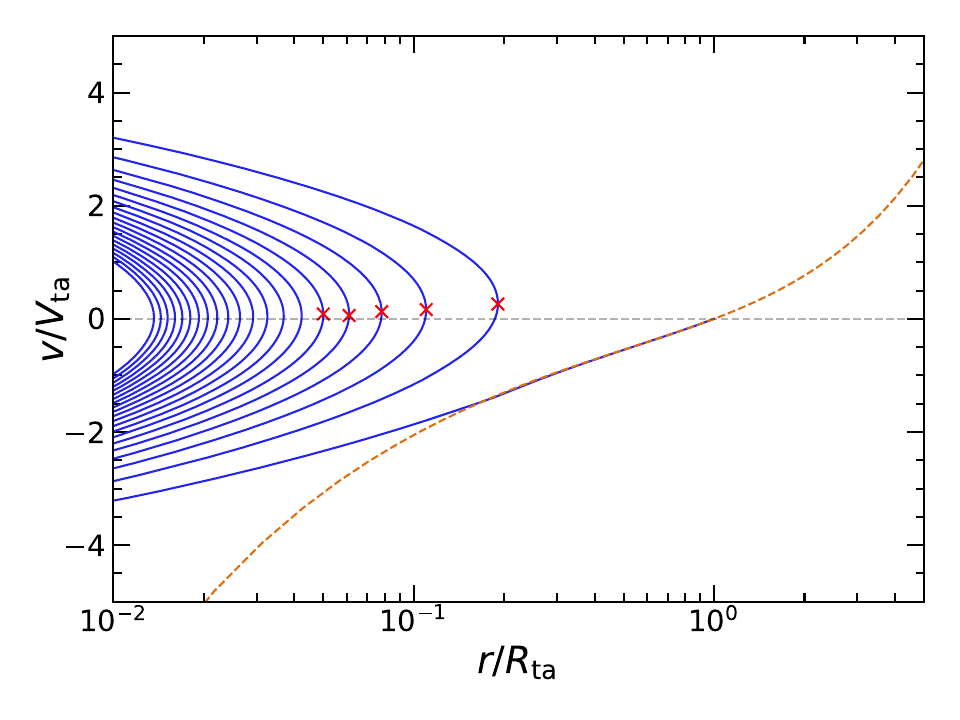}\\
    \includegraphics[width=0.49\textwidth]{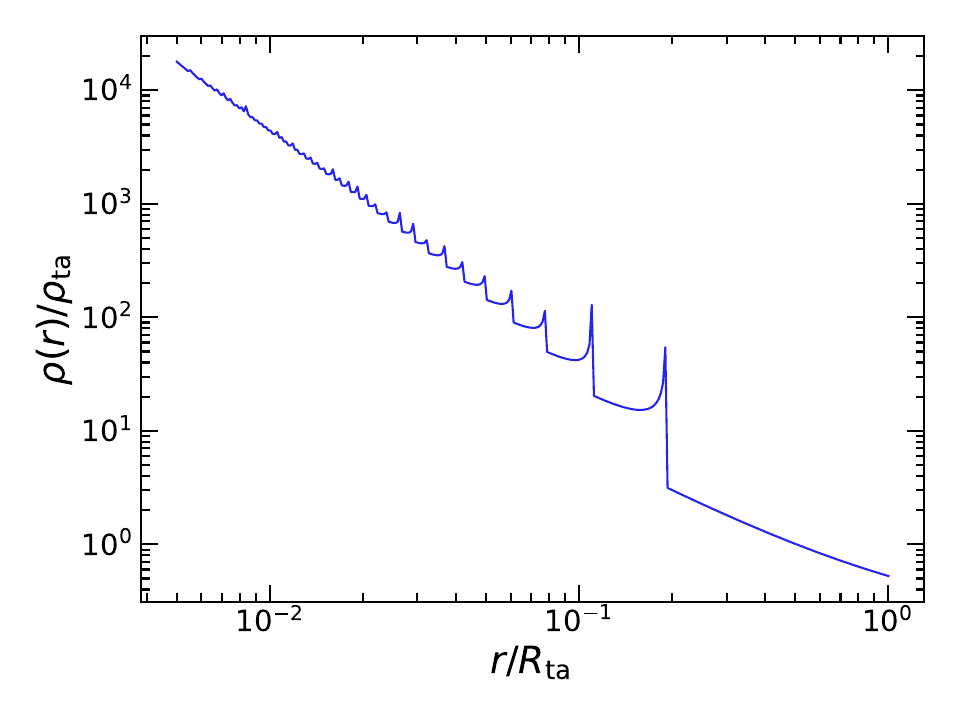}%
    \includegraphics[width=0.49\textwidth]{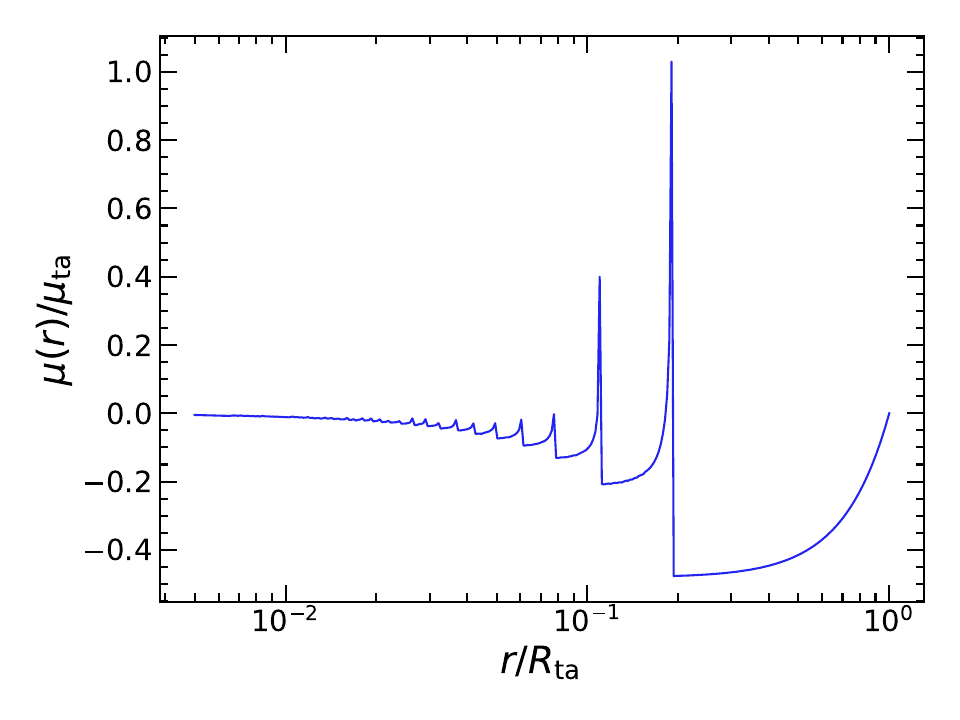}
    \caption{The dynamical structure of a halo formed from a $\epsilon=0.3$ peak. \textbf{Top left}: distance evolution of a shell, normalized by its own turnaround radius and time. 
    \textbf{Top right}: the phase space structure of the halo at a given time. The red crosses mark the locations of the first few caustics. The orange dashed curve shows the classical spherical collapse solution. \textbf{Bottom left}: the density profile normalized by the average density within the turnaround radius. Note some irregularities in the pattern of the caustics (density spikes) are due to the finite sampling of the profile when plotting.
    \textbf{Bottom right}: the mass flow rate profile (normalized by $\mu_{\rm ta}\equiv M_{\rm ta}/t$). Caustics corresponding to those in the density profile are observed in the outer halo, while the rate in the inner halo approaches zero.
    }\label{fig:ss_solution}
\end{figure}

Figure~\ref{fig:ss_solution} shows an example solution to \eqref{eq:SS_EoM}, confirming the qualitative behavior about halo collapse discussed above. Note the top panels can be interpreted both as specifying the distribution of shells with different $M_\ri$ at a fixed time $t$, or as the orbit of a given shell over time $t$. The apocentric distance as well as the orbital period of a shell shrinks over time due to the growing potential, while stabilizing in the inner halo where the potential is already deep. The decaying orbit causes the shell to spiral-in towards the inner halo in phase space. Note the enclosed mass of a shell decreases during infall in the shell-crossing regime, resulting in a lower kinetic energy compared with the classical spherical collapse model.

Because later-collapsing shells have a larger turnaround radius, $r_{\rm ta}$, their orbits can reach larger apocentric distances than earlier ones, causing the apoapses in the particle snapshot (the top right panel) to be located in the outflowing part of the phase-space trail (or equivalently, where $\dot{x}=0$). Projected into a density profile, the phase-space apoapses show up as caustics with diverging densities. In recent literature, the location of the outermost (or first) caustic is also referred to as the splashback radius.

To study the depletion radius, we now extend the self-similar model to derive the mass flow rate profile for the first time.
The mass flow rate can be found from the time derivative of Equation~\eqref{eq:Mprof} as
\begin{align}
\mu(r,t)&\equiv -\frac{\partial M(r,t)}{\partial t}\label{eq:mu_def}\\
&=\frac{M(r,t)}{t}(\frac{2}{3}+\frac{2}{9\epsilon})[\gamma(x)-\frac{1}{\epsilon+1/3}].\label{eq:MFR}
\end{align} where we define outflow as positive, and $\gamma\equiv\dfrac{\ud\ln\mathcal{M}(x)}{\ud \ln x}$. An alternative expression of the mass flow rate in terms of the velocity profile is provided in Appendix~\ref{app:practical_eqs}.


Note $\dfrac{\partial \ln M}{\partial \ln t}=\dfrac{\mu}{M/t}$, so the above equation can be rewritten as the logarithmic mass growth rate
\begin{equation}
s\equiv \frac{\partial\ln M(r,t)}{\partial \ln t}=(\frac{2}{3}+\frac{2}{9\epsilon})[\frac{1}{\epsilon+1/3}-\gamma(x)].
\end{equation}
In the inner halo, the mass profile evolves slowly and action is conserved. Realizing this, \citet{FG84} analytically derived that $\gamma=3/(1+3\epsilon)\leq 1$ and $s=0$ when $\epsilon\geq 2/3$, while $\gamma=1$ and $s>0$ when $\epsilon <2/3$. In the former case (steep peak), the inner halo is dominated by early collapsed shells and reaches equilibrium. In the latter case, the entire halo grows globally. Note the mass flow rate always approaches zeros in the center due to the additional $M(r,t)/t$ factor.

The caustics also show up in the mass flow rate profile at the same location as in the density profile, where the mass flow rate jumps from a negative value to a positive value. As a result, in the self-similar model, the depletion radius, defined at the maximum mass infall rate location, coincides with the splashback radius (see Figure~\ref{fig:ss_solution} for an example). However, in real halos, the two diverge (see section~\ref{sec:boundary_dm} for more discussions), and  theories beyond the spherical self-similar model are needed to understand this divergence.
\subsection{Non-spherical models}\label{sec:aspherical}
In the spherical solutions above, the particles move on purely radial orbits. However, in real $N$-body simulations, even if the initial conditions are spherically symmetric, radial orbit instabilities can make subsequent evolution of the halo non-spherical~\citep{Vogelsberger11}. Physical conditions such as the tidal torque exerted by the surrounding largescale environment can also act to generate non-radial motions. \cite{ZukinBertschinger10} showed that the angular momentum of particles generated by the tidal field can have an important influence on the inner density slope, as the pericenter of a particle is determined by the angular momentum. In the self-similar model, an angular momentum term can be specified as in Equation~\ref{eq:SS_EoM_j}. \cite{Nusser01} also studied the spherical collapse with non-zero angular momentum acquired at turnaround phase of particles.

\begin{figure}
    \centering
    \includegraphics[width=\linewidth]{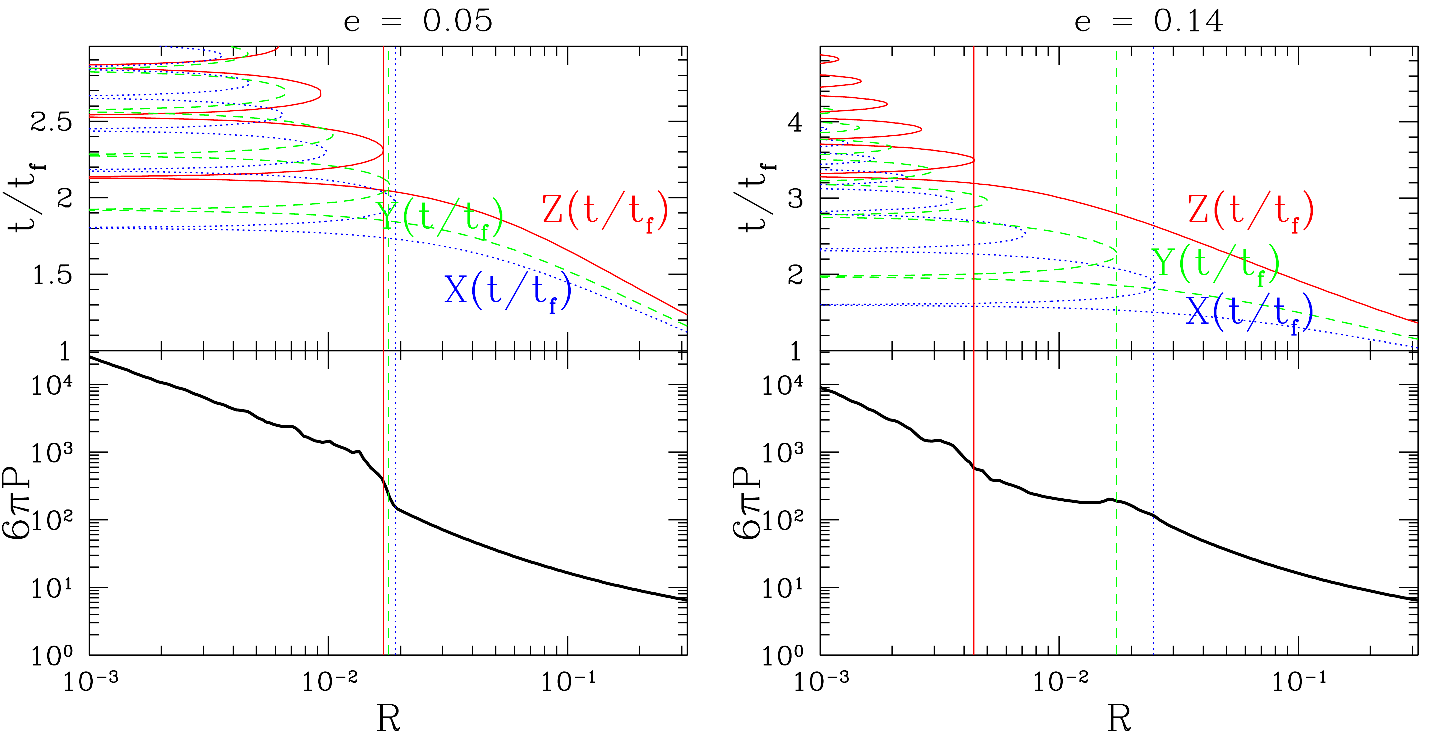}
    \caption{Self-similar collapse model of a tri-axial halo. The left and right panels show results for an ellipticity of $e=0.05$ and $e=0.14$, respectively, with the ellipticity describing the relative difference between the lengths of the major and minor axis. For each halo, the top panel shows the orbits of particles along the minor (X), middle (Y) and major (Z) axis in scaled coordinates, and the bottom panel shows the spherically averaged density profile of the halo. Because particles collapse at different times and splashback at different distances along different directions, the caustic features are smeared out in the resulting density profile. Figure reproduced from~\citet{LithwickDalal11}.}
    \label{fig:EllipCollapse}
\end{figure}

\cite{Ryden93} calculated self-similar collapse in the axisymmetric case. \cite{LithwickDalal11} generalized the self-similar collapse model to aspherical halos by considering initial perturbations that are separable in the radial and angular dependence while keeping the self-similar form in the radial component. In particular, they considered initial perturbations described by a triaxial ellipsoid. In this case, particles along different axis collapse at different paces, with the minor axis particles collapsing first and having the largest splashback radius, while those along the major axis splashback later and at a smaller radius (see Figure~\ref{fig:EllipCollapse}). This leads to a smearing of the splashback feature in the density profile, making the connection between the steepest slope and the apoapse of particles ambiguous. Note in this case, the depletion radius could also differ from the steepest slope radius.

\subsection{Generalization to realistic halo profiles and cosmology}
The self-similar solutions are restricted to power-law initial perturbation profiles in the EdS universe, both of which are scale free. 
The resulting mass profile has an inner slope $\gamma\leq 1$ corresponding to a density slope steeper than -2, differing from the inner slope of $-1$ expected for NFW halos. To gain insights into the formation of halo profiles and the boundaries of halos in more realistic conditions, it is necessary to develop more general spherical collapse models beyond the self-similar ones.

\citet{Adhikari14} built a toy model to derive the property of the splashback radius by evolving the orbit of a single shell falling into a growing NFW halo.
A more complete model is developed in \citet{Shi16}, in which the halo profile is calculated self-consistently from the trajectories of the shells rather than assumed to be NFW. Self-similarity of the halo mass profile is still assumed, $M(r, t)=M_{\rm hrta}(t)f(R/R_{\rm hrta})$, with the scales specified by the radius and mass at half of the turnaround radius, $R_{\rm hrta}$ and $M_{\rm hrta}$, which provide the scales of the virialized part of the mass profile. The mass accretion history of the halo is assumed to be a power law of the scale factor, $M_{\rm vir}\propto a^s$, and the trajectories of shells accreted at different time are solved simultaneously. \citet{Shi23} further extended this approach, referred to as the iterative mean field approach, to arbitrary initial perturbations or mass accretion histories. Because the dynamical timescale is usually shorter than the variation timescale of the mass accretion rate or the background cosmology, the mass profiles at each cosmic time can quickly converge without assuming self-similarity over time or mass.

These models provide direct predictions on the location of the splashback radius, which depends on both cosmology and the mass accretion rate. However, a direct comparison of the model predictions with simulations can be complicated because of the difference between the accretion rate evolution for a realistic halo and the constant accretion rate assumed by these models.

Beyond the splashbcack radius, the models of \citet{Shi16,Shi23} also shed light on the origin of the halo profile, as a consequence of the initial perturbation profile. A unique mapping between the mass accretion history, the profile shape and the initial perturbation profile can also be obtained. To explain realistic NFW profiles, however, aspherical collapse and angular momentum also need to be further considered.

\section{Boundary features in different tracers and observational measurements}~\label{sec:boundary}
\subsection{Dark matter}\label{sec:boundary_dm}
According to the secondary infall model, various features can be found around the transition between the single stream and multi-stream regions, giving rise to different characterizations of a halo boundary there. 
\subsubsection{Splashback radius}
In the idealized spherical collapse model, density caustics exist at the apocenter radii as in Figure~\ref{fig:ss_solution}, where the density rises sharply to infinity. In real halos, these caustic features get smoothed out due to various deviations from the model assumption, including deviation from spherical symmetry and from mergers. However, it is still possible to look for the caustics by identifying the radius where the density drops sharply over radius. \citet{DK14} first noticed that the radius where the logarithmic slope of the density profile is the steepest, later named as the splashback radius~\citep{Adhikari14,More15}, depends on the accretion rate of the halo, in line with expectations of the secondary infall model. 

Interpreting the splashback radius as the first apocentric radius after infall, \citet{SPARTA} developed the \textsc{sparta} code to identify the splashback of individual particles. The resulting radii show a wide spread in a given halo, such that the splashback process happens over a region rather than at a single radius. The steepest slope radius is on average close to the 75th percentile in the distribution of individual particle splashback.

 The splashback feature can be anisotropic. \cite{shellfish} developed the \textsc{shellfish} code to identify the splashback radius along different directions and connect them smoothly to form an anisotropic splashback shell. The effective splashback radius enclosing the same volume as the splashback shell is close to the 87th percentile~\citep{SPARTA2} in the distribution of the individual particle splashback radius, $r_{\rm sp,87}$.  The steepening feature along the halo major axis~\citep{Okumura18} or along the filament direction~\citep{Contigiani21} is found to be weaker compared to that along the minor axis. \citet{Huiyuan22} studied the density profiles along different directions with respect to the local tidal field. By associating the local minima in the gradient of the density profile with the first three caustics in spherical collapse, they found that the first caustic is always the most prominent feature along the minor axis of the tidal field, while that along the major axis (corresponding to the filament direction) could represent a different caustic, such that care must be taken when interpreting the spherically averaged density slopes.  

Some works utilized the velocity profile to define the splashback boundary. \citet{Deason20} defined a boundary at the steepest slope location in the velocity profile and found that it corresponds closely to that in the density profile~\citep[see also][]{Bose21,Pizzardo24,Sen26}. The splashback feature has also been identified in more complex kinematic statistics, including the momentum correlation function~\citep{Okumura18} and the vorticity profile~\citep{Luo23}. 

\citet{More15} provided a fitting formula for the dependence of the splashback radius on the mass accretion rate as 
\begin{equation}
    \frac{R_{\rm sp}}{R_{\rm 200m}}=0.54[1+0.53\Omega_m(z)](1+1.36e^{-\Gamma/3.04}).\label{eq:RspfitMore}
\end{equation} This fitting formula extends upon the original fitting function of \citet{DK14}, with a redshift dependence parameterized through the cosmological parameter at each redshift. Such a parameterization is motivated by theoretical expectations, although the results are only calibrated with a single simulation. Given that halos of different masses have different mass accretion rates on average, Equation~\eqref{eq:RspfitMore} can be converted to predict the average splashback radius for halos of different masses. \citet{More15} found that this mass dependence varies weakly with redshift when expressed as a function of the peak height, which can be fitted as
\begin{equation}
    \frac{R_{\rm sp}}{R_{\rm 200m}}=0.81(1+0.97e^{-\nu/2.44}).
\end{equation} While these results are calibrated with the splashback radius defined at the steepest slope location, alternative fitting results for the radius determined using other techniques can be found in \citet{SPARTA2,shellfish}. 

\subsubsection{Depletion radius}
An alternative and more general approach for defining a boundary around the halo is to study the time evolution of the halo density profile. As a halo grows, it also drains its environment, causing the surrounding density to drop over time. The evolution of the density is governed by the continuity equation,
\begin{equation}
    \frac{\partial \rho}{\partial t}=-\frac{1}{4\pi r^2}\frac{\partial \mu}{\partial r},
\end{equation}
where $\mu = 4\pi r^2\rho v_r$ is the mass flow rate (see also Equation~\ref{eq:mu_def}).
The transition between the growing density and the decreasing density happens where the mass flow rate towards the halo, $-\mu$, is the maximum. \citet{FH21} defined this radius as the depletion radius. As illustrated in Figure~\ref{fig:ss_solution}, in the ideal spherical collapse model, this depletion radius is also located at the first caustic. It is generated when the infalling stream confronts the back-splashing particles, causing the net mass inflow rate to decline. However, in real halos it is found that the depletion radius is on average two times the virial radius~\citep{FH21,Gao23}, and corresponds to an outer edge of the splashback region enclosing a highly complete fraction of splashback particles. By contrast, the steepest slope location corresponding to the classical splashback radius is larger than the virial radius only by $\sim 25\%$. Nevertheless, the definition of the depletion radius is independent from the splashback model, and can arise through whatever mechanism that drives the evolution of the density profile. For example, the depletion radius can be equally well-defined in the gas profile, for which the splashback process is less relevant than the accretion shock.

The depletion radius is also shown to depend sensitively on the mass accretion rate. \citet{Jiale} provided a fitting formula as
\begin{equation}
\frac{R_{\rm id}}{R_{\rm vir}}=1.35+0.42\ln\Omega_{\rm m}(z)+[1.12-0.51\ln\Omega_{\rm m}(z)]\mathrm{e}^{-\Gamma/3}. \label{eq:ridrvir}
\end{equation}
Note the above equation is normalized by $R_{\rm vir}$ instead of $R_{\rm 200m}$. 
Major mergers play an important role in driving the redshift dependence of the depletion radius. Excluding halos that have experienced recent major mergers, the redshift dependence of the depletion radius can be largely removed, reducing Equation~\eqref{eq:ridrvir} to its high redshift counterpart ($\Omega_{\rm m}\approx1$). The accretion rate is found to be the primary driver of the depletion radius in the slow accretion regime (for $\Gamma<1$), while the long term history and the detailed accretion mode such as anisotropic infall are also important contributors in the fast accretion regime. The origins of these dependences remain to be investigated.

By contrast, the accretion rate dependence of the splashback radius is not sensitive to major mergers. When normalized by $R_{\rm vir}$ instead of $R_{\rm 200m}$, \citet{Jiale} found the relation to be insensitive to redshift either, with
\begin{equation}
\frac{R_{\rm sp}}{R_{\rm vir}} = 0.75 + 1.04 \mathrm{e}^{-\Gamma/3}. \label{eq:rsprvir}
\end{equation}
Combining Equation~\eqref{eq:rsprvir} with the smooth-accretion ($\Omega\approx1$) version of Equation~\eqref{eq:ridrvir},
the two dynamical radii can be approximately related by a constant offset of
\begin{equation}
    R_{\rm id}(\Gamma)\approx 0.54R_{\rm vir}+R_{\rm sp}(\Gamma),
\end{equation} applicable to smoothly accreting halos.

The depletion radius has been found to be well suited for serving as an optimal ``exclusion radius''  in the halo model of the large scale structure~\citep{ZhouHan23,ZhouHan25}, which we further discuss in section~\ref{sec:halomodel}.

\subsubsection{Phase space boundaries}\label{sec:phasespaceboundary}
Some recent works also propose to define the boundary of a halo in phase space rather than in configuration space alone. \citet{Aung21} decomposed the particles around a halo into an orbiting and an infalling component separated by the first pericenter passage after infall. According to this decomposition, they also defined an edge radius where the fraction of the orbiting particles drops to 1 percent. The edge radius is found to be approximately 1.6 times $r_{\rm sp,87}$, which is roughly the same location as the depletion radius~\citep{FH21}. Such an edge radius could also be observable through statistical decomposition of the galaxy velocity distribution around cluster halos~\citep{Tomooka20, Aung23}.  \citet{DiemerDecompose} extended the \textsc{sparta} code to decompose halo particles into the orbiting and infalling components. The density profile of the orbiting component was found to be well described by a smoothly truncated Einasto profile~\citep{DiemerDecompose2}, with a truncation radius that depends sensitively on the accretion rate of the halo~\citep{DiemerDecompose3}. \citet{Garcia23} developed an alternative algorithm for classifying the orbiting and infalling particles based on the accretion time and radial velocity of a particle. 

A natural extension of the orbiting-infalling decomposition is to decompose the halo into multiple stream components. \citet{Sugiura20} partitioned halo particles according to the number of apocenter passages since infall using cosmological simulations in the EdS universe, and compared their phase space distributions with the self-similar solution. They found that about 30\% of halos can be well described by the self-similar model. \citet{Enomoto23} and \citet{Enomoto24} further extended the analysis to $\Lambda$CDM as well as to warm dark matter simulations~\citep{Enomoto25} and found that the density profile for particles orbiting within each period can be well described by a double power-law form with an inner slope of $-1$ and an outer slope of $-8$. Summing up the double power law profiles from different accretion periods recovers the total density profile. On the other hand, the self-similar model, as well as its extensions~\citep{Shi16,LithwickDalal11}, generally predict an inner density profile slope of $-2$ or steeper, differing from the inner slope of $-1$ from each stream. Introducing a large angular momentum can partly alleviate the discrepancy by truncating the individual stream profile at a finite pericenter distance, thereby producing a flatter total profile. To match the individual stream profiles, it may be necessary to account for realistic distribution of angular momentum and deviations from spherical symmetry (see discussions and references in section~\ref{sec:aspherical}, or more realistic accretion histories~\citep{Lu06}. 
It remains a theoretical task to develop more accurate models for the stream structure through both analytical and simulation efforts. 

 \subsubsection{Observations}
As the boundaries of a halo are typically defined via characteristics in the density profile, they can be inferred through direct measurements of the profile. Inferring the virial mass or radius defined according to the virial density is a classical topic in weak lensing~\citep[e.g.,][]{Mandelbaum06,Han15Lensing,ClusterLensing,GapLensing} and dynamical modelling~\citep[e.g.,][]{ClusterDynamics,MWMassRev} studies. Extending these studies to larger scale, the outer boundaries could also be measured. By fitting the stacked lensing profile with a parametric density profile model of either \citep{DK14} or \citep{FH21}, multiple works have measured the splashback radius in the total matter profile~\citep{Umetsu17,Chang18,Contigiani19,Shin21,Fong22,Xu24,Giocoli24,Lesci26}. The depletion radius has also been probed observationally with weak lensing, though somewhat indirectly by identifying the location of the minimum bias around galaxy clusters, which is a consequence of the depletion process~\citet{Fong22}. 

These lensing measurements are largely consistent with theoretical expectations, though some inconsistencies have been observed especially when splitting the halo sample according to some secondary properties other than mass. For example, \citet{Fong22} found that lower-concentration clusters tend to have relatively larger splashback and depletion radii, consistent with the observational measurement of \citet{More16}. However, in simulations lower-concentration halos generally correspond to later forming ones with higher accretion rates, which tend to have a smaller splashback-to-virial ratio. These discrepancies may be associated with systematics in the sample selection or analysis method, as it is challenging to find an accurate proxy for the halo accretion rate and to accurately account for its distribution conditioned on a certain observational concentration proxy.


\subsection{Gas}
The collapse of a collisional gas component has also been investigated in the self-similar secondary infall framework in \citet{Bertschinger85} (for $\epsilon=1$) and recently extended in \citet{Shi16b} to arbitrary mass accretion rates, both in the EdS universe. 
\citet{Bertschinger85} considered gas collapse both with and without a black hole in the center, and both with and without a dark matter component. Here we focus on the model with both gas and dark matter and without a central black hole. Assuming the gas mass is negligible compared to that of the dark matter component, then the collapse of dark matter follows those described in section~\ref{sec:selfsimilar-sol}. The gas component also follows the same dynamics initially, until it confronts an accretion shock leading to a jump in its properties, after which it gets further slowed down by the pressure gradient and eventually comes to rest at the center. The post-shock gas dynamics are specified by the fluid equations,
\begin{align}
    &\frac{\ud \rho}{\ud t}=-\frac{\rho}{r^2}\frac{\partial}{\partial r}(r^2 v)\\
    &\frac{\ud v}{\ud t}=-\frac{GM}{r^2}-\frac{1}{\rho}\frac{\partial p}{\partial r}\\
    &\frac{\ud}{\ud t}(p\rho^{-\gamma_g})=0\\
    &\frac{\partial M_{\rm gas}}{\partial r}=4\pi r^2 \rho,
\end{align} where $\rho,p,v$ are the density, pressure and velocity of the gas, and $M_{\rm gas}$ and $M$ are the gas and total masses. The gas is assumed to be adiabatic with a ratio of specific heats $\gamma_g$. Self-similar solutions to these equations, as well as the location of the shock radius, can be found with the outer boundary conditions at the shock jump and the inner boundary conditions at the center. The solution depends on both the mass accretion rate (or equivalently the initial profile shape, $\epsilon$) and the adiabatic index, $\gamma_g$. Same as the splashback radius, the shock radius decreases with the mass accretion rate~\citep{Shi16b}. The shock radius is close to the splashback radius for $\gamma_g\approx 5/3$ but increases with $\gamma_g$. In this idealized model, the shock radius also corresponds to the gas depletion radius, as the gas inflow rate peaks right before reaching the shock boundary.

Despite the simple theoretical picture above, the gas boundaries in simulation appear much more complex. In addition to the accretion shock discussed above, another important channel for shock creation is through mergers. When another gaseous structure (halo or galaxy) falls into the inner part of a halo, the supersonic infall velocity can generate a bow shock ahead of it~\citep{Ryu03}. Once the substructure passes the pericenter, the bow shock can detach from the substructure, forming a run-away shock~\citep{Congyao19}. When the smooth accretion rate is low, the runaway shock can overtake the accretion shock, generating a secondary merger-accelerated accretion shock well outside the splashback radius~\citep{Congyao20,Congyao21}. 

\begin{figure}
\begin{center}
\includegraphics[width=0.6\textwidth]{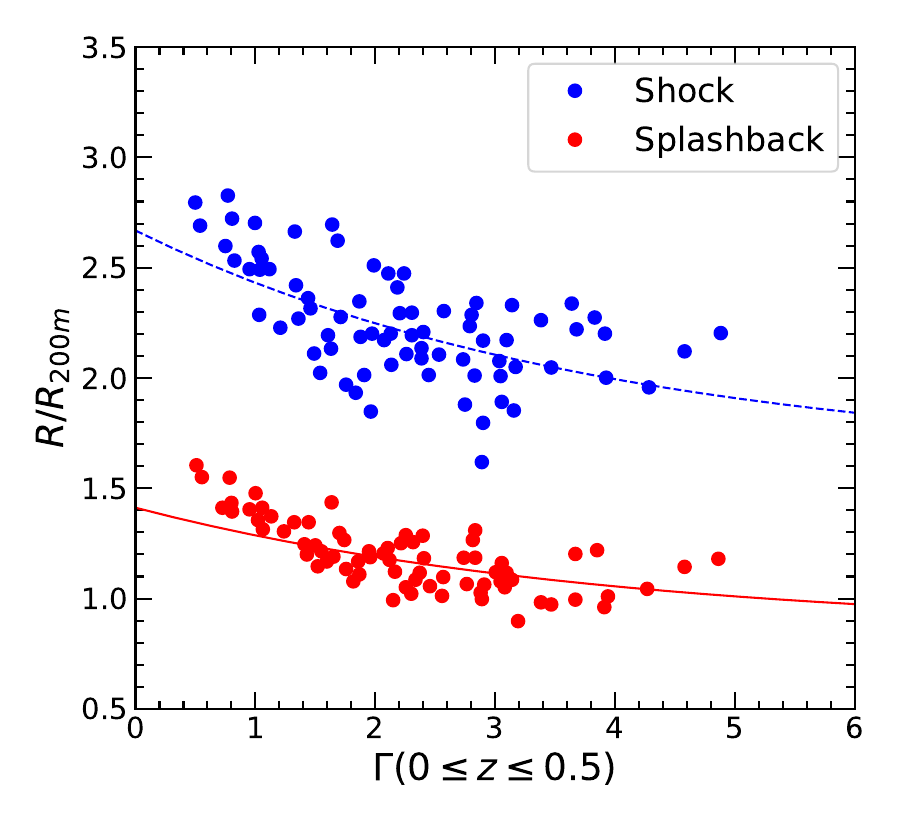}
\end{center}
\caption{The relation between the shock radius (blue dots) and the splashback radius (red dots), for clusters of different accretion rate, $\Gamma=\ud \log M_{\rm 200m}/\ud \log a$, in hydrodynamical simulations \citep{Aung21gas}. Both radii are rescaled by the cluster virial radius, $R_{\rm 200m}$. The shock radius is defined at the steepest slope location in the gas entropy profile. The red solid line shows the fitting result for the splashback radius from \citet{shellfish} for halos of peak height $\nu=3$. The blue dashed curve is a rescaled version of the black solid curve by a factor of 1.89.}\label{fig:shockrad}
\end{figure}

Simulation measurements of the shock radius result in different answers depending on the features adopted to identify the radius. Some studies have measured the steepest slope location in the gas density profile of galaxy clusters, and found it to be $\sim 10-20\%$ smaller than that in the dark matter profile~\citep{Oneil21, Aung21gas,Zhang25}. 
Alternative and probably more physical definitions of the shock radius can be constructed from other thermodynamic profiles of the gas. For example, \citet{Lau15} defined the location of the peak in the entropy profile as the shock radius, which is found to be on average $1.38$ times the splashback radius~\citep[see also][]{Zhang25}. \citet{Aung21gas} found that the steepest slope locations in the gas entropy, temperature, and pressure profiles are on average $\sim 1.8-1.9$ times the splashback radius defined in the dark matter density profile (see Figure~\ref{fig:shockrad}). \citet{Baxter21} found that the steepest slope location in the stacked Compton-y profile, which can be directly measured through the Sunyaev-Zel'dovich (SZ) effect, is close to the location of the maximum gas infall velocity, and is on average twice the splashback radius. \citet{Sen26} found that the offset between the gas shock radius and the dark matter splashback radius occurs predominantly along the void directions. The also pointed out that the offset tends to be small at high redshift, but develops following major merger events.

The dark matter splashback radius is found to be insensitive to baryonic physics, with less than $5\%$ difference between hydrodynamical and dark matter only simulations~\citep{Oneil21,Towler24}, while the gas radii can be more affected, with $\sim 10\%$ differences~\citep{Towler24} across different baryonic models in the FLAMINGO simulations~\citep{FLAMINGO}. 
The shock radius, irrespective of the definition, is often found to decreases with increasing mass accretion rate~\citep{Oneil21,Aung21gas,Zhang25}, in line with the dependence of the splashback radius. However, some contradicting results also exist in the literature, which may be due to the technical details in computing the various profiles. For example, \citet{Towler24} found the steepest slope location in the pressure profile  is close to the splashback radius rather than the minimum in the entropy gradient, in contradiction to \citet{Aung21gas}. They also found that the minimum in the gas density and pressure gradients are sensitive to halo mass but not to accretion rate. Further explorations are needed to fully understand the origin of the different gas radii and their connections to the splashback radius.

Observational, \citet{Anbajagane22} identified two steepening features in the Compton-y profile of SZ-selected clusters, one located around $R_{\rm 200m}$, while the other about $4.5$ times as large. \citet{Anbajagane24} extended the analysis using a larger cluster sample, and identified the first steepening as the shock radius. This radius is about 1.17 times the splashback radius identified from the steepening of the galaxy counting and weak lensing measured mass profiles for the same sample of clusters~\citep{Shin19}, largely consistent with theoretical expectations. 

\subsection{Galaxies, subhalos and halo stars}
Galaxies, subhalos and halo stars are to a large extent collisionless tracers moving in the potential of the host halo, and thus are expected to exhibit similar boundary features in their phase space distributions. Studies using numerical simulations generally show that the splashback features in the profiles of these tracers are located close to that in the total dark matter profile, though at slightly larger or smaller radii. For example, using the IllustrisTNG simulation and selecting galaxies to be luminous subhalos above $10^9\msun$, \citet{Oneil21} found that the splashback radius identified in the galaxy number profile is close to that in the dark matter density for halos above $10^{13.5}\msun$, but smaller by $\sim 12\%$ in $10^{13}$ to $10^{13.5}\msun$ halos. Observational measurements using cluster-galaxy cross correlation also found a $\sim 20\%$ smaller splashback radius than expectations from the corresponding subhalo population in dark matter-only simulations~\citep{More16,Chang18}. On the other hand, the measurements from cluster-galaxy correlation tend to be smaller than those from weak lensing for the same sample of clusters~\citep{Chang18,Lesci26}. These results, however, could be affected by systematics in the cluster finding~\citep{BuschWhite17,Murata20} or in the galaxy selection~\citep{Oneil22}. Measurements using X-ray and Sunyaev-Zel'dovich (SZ) selected clusters generally show better agreement with simulation expectations~\citep{Zurcher19,Shin19,Rana23} as well as more consistent results between galaxy counts and weak lensing profiles~\citep{Shin21}.

\citet{Deason21} showed that the splashback radius in the stellar mass profile is close to that in the dark matter profile in simulated clusters~\citep[see also][]{Zhang25}, while care must be taken on the influence of substructures. Analogously, \citet{Walker26} studied the orbiting-infalling decomposition of simulated cluster halos for both the dark matter and the stellar components. They found that the truncation radius in the orbiting component of the stars coincides with that of dark matter, and thus provides an observational proxy for its dark matter counterpart. As the stellar mass profile is much steeper than the dark matter profile in the outskirts, the splashback radius in the stellar profile is beyond the reach of existing observations of the intracluster light~\citep[e.g.,][]{Wenting19,Zhangyuanyuan19}, but can be accessible in future observations using Vera C. Rubin Observatory~\citep{LSST}, the Nancy Grace Roman Space Telescope~\citep{Roman}, and Euclid~\citep{EUCLID}.  

Using simulations of Milky-Way-like halos, \citet{Deason20} found that in general the outermost two caustics can be identified in the dark matter distribution, in both the density and velocity profiles. Meanwhile, the distribution of stars is more concentrated than dark matter and the second caustic is the most prominent feature. For subhalos, the most prominent caustic feature is the outermost one, while for luminous galaxies, it is the second caustic. These detected caustics, on average, are well aligned with the corresponding ones in the dark matter distribution. Detailed differences between the locations of the minima in the density slopes of dark matter and stellar components can be understood as dark matter and stars are deposited at different rate to the halo during mergers~\citep{Genina23}. Using observed velocities of nearby dwarf galaxies, \citet{Deason20} estimated a tentative splashback radius of the MW halo at $r_{\rm sp}=292\pm61{\,\rm kpc}$, which is about $0.8r_{\rm 200m}$ and almost the same as the top-hat virial radius, $r_{\rm vir}$~\citep[e.g.,][]{Li20}. Using a similar dataset, \citet{LiHan21} performed the first dynamical measurement of the depletion radius by identifying the maximum infall velocity location around the MW at $r_{\rm id}=559\pm107{\,\rm kpc}$, and the turnaround radius at $r_{\rm ta}=839\pm121{\,\rm kpc}$.

Dynamical friction can play an important role in determining the splashback radius of satellite galaxies. As dynamical friction is more important for more massive subhalos, the splashback radius for massive satellites is expected to be smaller than that of low-mass ones. Indeed, \citet{Adhikari16} detected a systematic decrease of the steepest slope radius with satellite magnitude around a sample of SDSS galaxy clusters. However, there are still debates on how much dynamical friction can explain the systematic decrease. For example, some works do not find a sensitive dependence of the observed splashback radius on galaxy magnitudes~\citep{More16,Murata20}. Experimenting with idealized simulations, \citet{OShea25} argues that dynamical friction cannot be responsible for the observed systematics in clusters more massive than $10^{14}\msun$.   

\section{Statistics of new boundaries and applications in structure formation}
The various dynamical boundaries, including the splashback radius, depletion radius, and turnaround radius, are by construction immune from the pseudo evolution problem, as their definitions do not rely on the assumed co-evolution model of the halo and the background universe. More importantly, the new boundaries can also bring new insights into the understanding of halo evolution and structure formation. The application of the splashback radius as a probe of the mass accretion rate of the halo covered in section~\ref{sec:boundary_dm} provides a good example of the power of the new boundaries. In addition, a few recent works have highlighted the power of the depletion radius in serving as a physical probe for halo growth and as an optimal boundary for the halo model of the largescale structure. We summarize these progresses below.

\subsection{Evolution of the halo profiles}

\begin{figure}
    \centering
    \includegraphics[width=\linewidth]{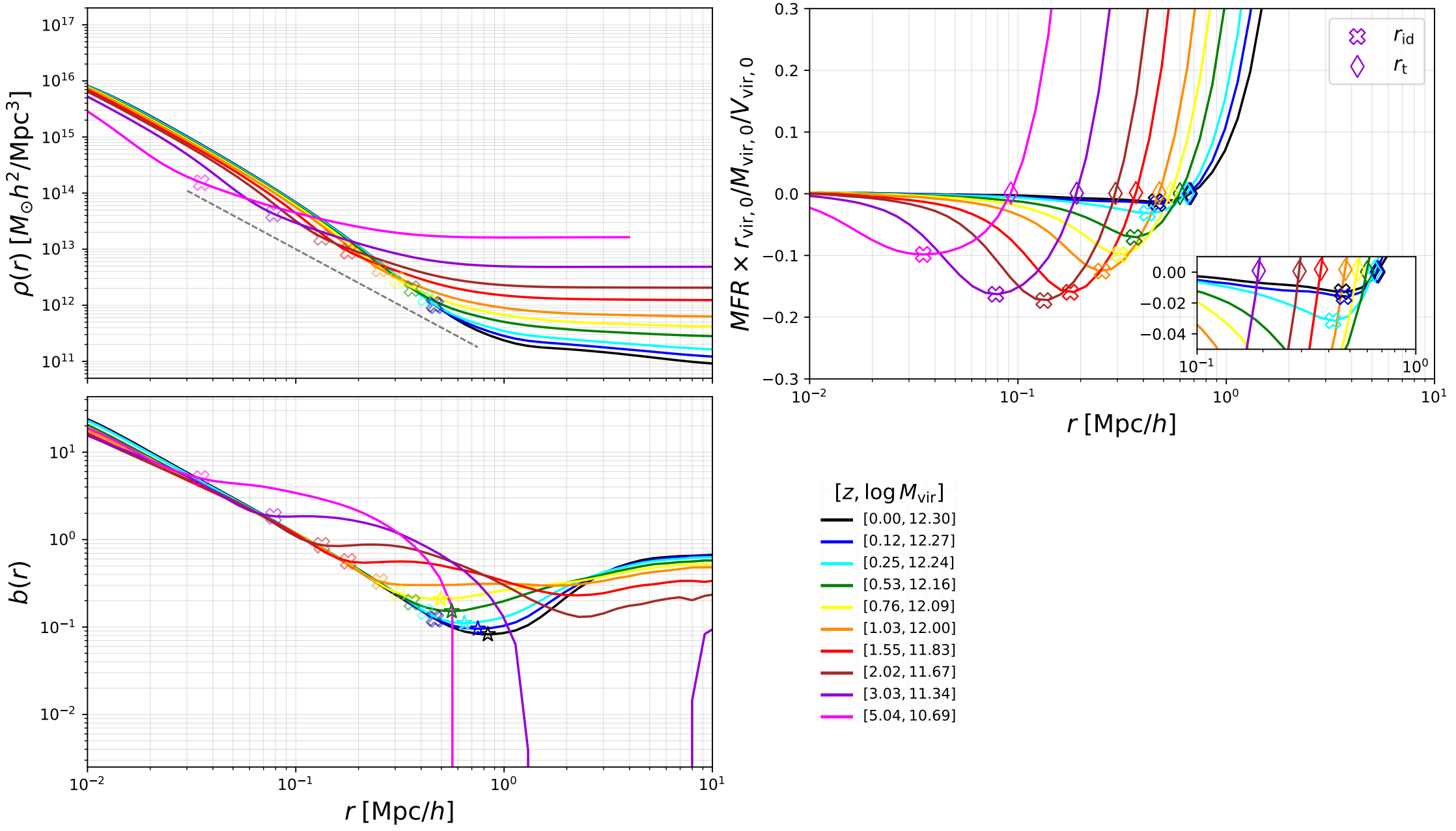}
    \caption{Evolution of the density (top-left), mass flow rate (top-right) and bias profiles (bottom-left) around a sample of galactic size halos in a cosmological simulation. The halos are selected at $z=0$ and traced back in time, with their average mass (in unit of $\msunh$) at each redshift shown in the legend at the bottom-right corner. For each profile, the cross and diamond with the corresponding color mark the depletion and turnaround radius at the corresponding redshift, respectively, while the stars in the bias profiles mark the characteristic depletion radii. The dashed curve in the density panel show a reference line of $\rho\propto r^{-2}$. The mass flow rate profiles are normalized by the virial quantities at $z=0$, and the inset panel shows a zoom-in of the late-time profiles around the depletion radii. All quantities shown are in physical rather than comoving coordinates. Figure adapted from \citet{Gao23}.}
    \label{fig:prof_evo}
\end{figure}

\citet{Gao23} showed that the depletion radius can provide a very physical and intuitive description of the evolution of the halo profiles. As shown in the top-left panel of Figure~\ref{fig:prof_evo}, the density profile around a halo grows in the inner region but decreases in the outer part, with the depletion radius separating the two regions by definition. Note that cosmic expansion also causes the physical density to decrease. To separate this effect out, it is useful to rescale the density profile by the average density profile of the universe. This is effectively achieved by studying the bias profile,
\begin{equation}
    b(r|h)=\frac{\xi_{\rm hm}(r)}{\xi_{\rm mm}(r)}.
\end{equation} Here the halo-matter correlation, $\xi_{\rm hm}$, represents the average overdensity profile around the halo sample, while the matter-matter correlation, $\xi_{\rm mm}$, represents that around a random matter particle in the universe. As shown in the bottom-left panel of Figure~\ref{fig:prof_evo}, the bias profile evolves by depleting material outside the depletion radius, gradually forming a trough around the boundary. As a result, the depletion process provides an intuitive interpretation of the existence of a minimum in the nonlinear bias profile around the halo. The location of the bias minimum was also named as the characteristic depletion radius in \citet{FH21} and provides a proxy for observing the depletion radius through measurement of the density profile~\citep{Fong22}.

The depletion process is \emph{self-regulated}, happening at an accelerating rate initially, followed by a decelerating phase as the surrounding material becomes sparse. This is shown in the top-right panel of Figure~\ref{fig:prof_evo}, in which the mass inflow rate ($-MFR$) at the depletion radius first increases over time and then decreases. \citet{Gao23} pointed out that the transition epoch between the two depletion phases corresponds well to the well-known transition between the fast and slow accretion phases governing the growth of the inner halo~\citep{Zhao03}, and provides a natural and physical explanation for the origin of the latter. Such a transition is also clearly visible in the relative variation rate of the inner density profile in Figure~\ref{fig:prof_evo}.

\subsection{Halo distribution and large scale structure}

One major application of dark matter halos is for constructing models of the largescale structure. The halo model approach to structure formation essentially assumes that the cosmic density field, $\rho(\textbf{\textit{x}})$, is equivalent to a halo field, $n(\textbf{\textit{x}})$, convolved by the internal structure of each halo, $\rho_h(\textbf{\textit{x}})$, with
\begin{align}
\rho(\textbf{\textit{x}})&= \sum_i \rho_{\rm h}(\textbf{\textit{x}}-\textbf{\textit{x}}_i)\nonumber \\
&= \sum_m n(\textbf{\textit{x}}|m)\circledast \rho_{\rm h}(\textbf{\textit{x}}|m),
\label{eq:matter_field}
\end{align}
where $\circledast$ represents the convolution operator. The summation in the second equation runs over halos of different properties in general, while for simplicity, we represent them with a single mass label, $m$, here.

As a result, the statistical properties of the halos as functions of their structural parameters fully specify the statistics of the cosmic density field. With the mass or size of a halo as a primary structural parameter, the abundance, clustering, and the density profile as functions of halo mass form three key ingredients for a basic halo model.


\subsubsection{Halo population statistics}\label{sec:population_stat}
The definition of a halo boundary can affect the statistics of halos in two aspects. First, a different boundary definition results in a different mass label for the same object. Second, a larger boundary means some halos that were defined as isolated become merged as one object if they overlap on the new boundary. The second effect mostly reduces the number of low mass objects that are close to larger objects, thus reducing the halo mass function at the low mass end if the same mass label is used for counting halos~\citep{Garcia19}. The halo bias at the low mass end will also be reduced, as the removed halos are the more biased ones due to their proximity to larger objects.

Despite this, the general formulas describing the classical halos are found to be still applicable to halos following the new boundary definitions. For example, \citet{ZhouHan23} found that the Sheth-Tormen formula can also describe the mass function of halos isolated on the depletion radius in the concordance cosmology, though with different parameters.  Besides, the bias fitting formula developed for classical halos~\citep[][]{jing1998accurate} can also be applied to the depletion radius based halo sample. 
\citet{Garcia23} also found that both the mass function and the bias for dynamical halos, defined to include only the orbiting particles, can be described in the excursion set formalism. The halo mass function at different redshift and cosmology can, however, deviate from a universal form to varying degrees depending on the mass definition~\citep[e.g.,][]{Despali16, Diemer20Universal}, by as much as $20\%-500\%$ at the high $\nu$ end. \citet{Diemer20Universal} investigated the universality of the halo mass functions according to the splashback-radius-based mass definitions measured with the \textsc{sparta} code, and found tighter universality than that in the virial mass functions for $z\leq 2$ as well as for self-similar cosmologies. 


The density profile of a halo can be explicitly measured inside a given halo boundary. However, it should be kept in mind that real halos are not abruptly truncated at a given boundary. Asphericity of the mass distribution, as well as ongoing mergers, can contribute to a continuous mass distribution outside any given boundary, which needs to be properly accounted for when using halos to reconstruct the matter field. Classical works usually assume the halo profile follows the Navarro-Frenk-White~\citep[NFW,][]{navarro1995simulations,navarro1996structure,navarro1997universal} profile with some truncation in the outskirts to avoid divergence in total mass. \citet{ZhouHan25} realized that different boundary definitions require different density profiles for reconstructing the matter field. As the matter field is obtained by convolving the halo field with the halo profiles, the optimal profiles can be derived knowing the matter and halo fields~\citep[see also][]{ADM1,ADM2}, essentially inverting Equation~\ref{eq:matter_field} in Fourier space to obtain $\rho_h(x|m)$. Following this approach, \citet{ZhouHan25} found that the optimal profile required for halos defined by the depletion radius can be well described by the Einasto profile~\citep{einasto1965construction, Merritt06, Navarro04, Navarro10}, 
\begin{equation}
    \rho_{\rm EIN}(r) = \rho_{\rm s} {\rm exp}\left( -\frac{2}{\alpha}\left[ \left(\frac{r}{r_{\rm s}}\right)^\alpha-1 \right]\right),
    \label{eq:EIN}
\end{equation} where $\rho_{\rm s}$ and $r_{\rm s}$ are the scale density and scale radius, while $\alpha$ is tightly correlated with the peak height parameter~\citep{Gao08}. A larger boundary requires a more extended profile. 

\subsubsection{Large scale structure model}~\label{sec:halomodel}
A physical halo boundary definition can also resolve the difficulty in modeling the clustering of halo and matter on the intermediate scale, i.e., solving the halo exclusion problem. This difficulty arises mainly from two aspects. First of all, the classical boundary following the virial definition does not correspond explicitly to where halos end. As a result, there is  large ambiguity in  how to model the density profile outside the virial radius, and where to truncate the density profile. Secondly, the clustering of halos on the intermediate scale follows neither the linear nor the non-linear matter clustering, necessitate the introduction of a so-called non-linear bias, which has to be modelled empirically~\citep[e.g.,][]{hayashi2008understanding,tinker2005mass,van2013cosmological, HMCODE15, Philcox20EHM}.

\citet{Garcia21} constructed a model for the halo-mass correlation function, $\xi_{\rm hm}$, that explicitly incorporates halo exclusion and allows for a redefinition of the halo boundary in a flexible way. They found that an optimal exclusion radius that best describes the nonlinear clustering lies around the minimum in $r^2\xi_{\rm hm}$ for cluster halos, which is close to 1.3 times $r_{\rm sp,87}$. 
Intriguingly, \citet{FH21} showed that this optimal exclusion radius is very close to the depletion radius.  Coupled with the Einasto profile (see section~\ref{sec:population_stat}), the depletion radius thus solves the ambiguity in defining the halo exclusion radius and the associated density profile.

Defining halos using the depletion radius, \citet{ZhouHan23} found that the halo-halo correlation function takes on a universal form which becomes self-similar on the nonlinear scale, all the way down to the sum of the depletion radii where two halos exclude each other. Specifically,
\begin{equation}
    1+\xi_{\rm hh}^{\rm 2h}(r|m_1,m_2)=[1+b(m_1)b(m_2) \hat{\xi}_{\rm hh}(r)]\mathrm{H}(r-r_{\rm t}),\label{eq:xi_hh_fit}
\end{equation} where $\mathrm{H}(x)$ is the Heaviside step function, and $r_{\rm t}=R_{\rm id}(m_1)+R_{\rm id}(m_2)$ is the halo exclusion distance. The unit halo-halo correlation function takes on the form
\begin{equation}
    \hat{\xi}_{\rm hh}(r)=\left(\frac{r}{r_0}\right)^{-\gamma},\label{eq:xi_hh_star}
\end{equation} on intermediate scale and connects to the linear correlation on large scale.
In this form, a single linear bias parameter, $b(m)$, is sufficient for describing the halo distribution across scales, and diffuse matter can also be accounted for self-consistently, thereby solving the challenge in modelling the non-linear halo clustering. Note the unit halo-halo correlation function only emerges for the depletion radius based halo catalog, while the correlation for the classical virial radius based halos are much more complicated and difficult to model. Convolving the halo-halo correlation with the Einasto profile, the matter distribution on the inter-halo scale can be accurately reproduced. 

\begin{figure}
    \centering
    \includegraphics[width=0.5\linewidth,height=233pt]{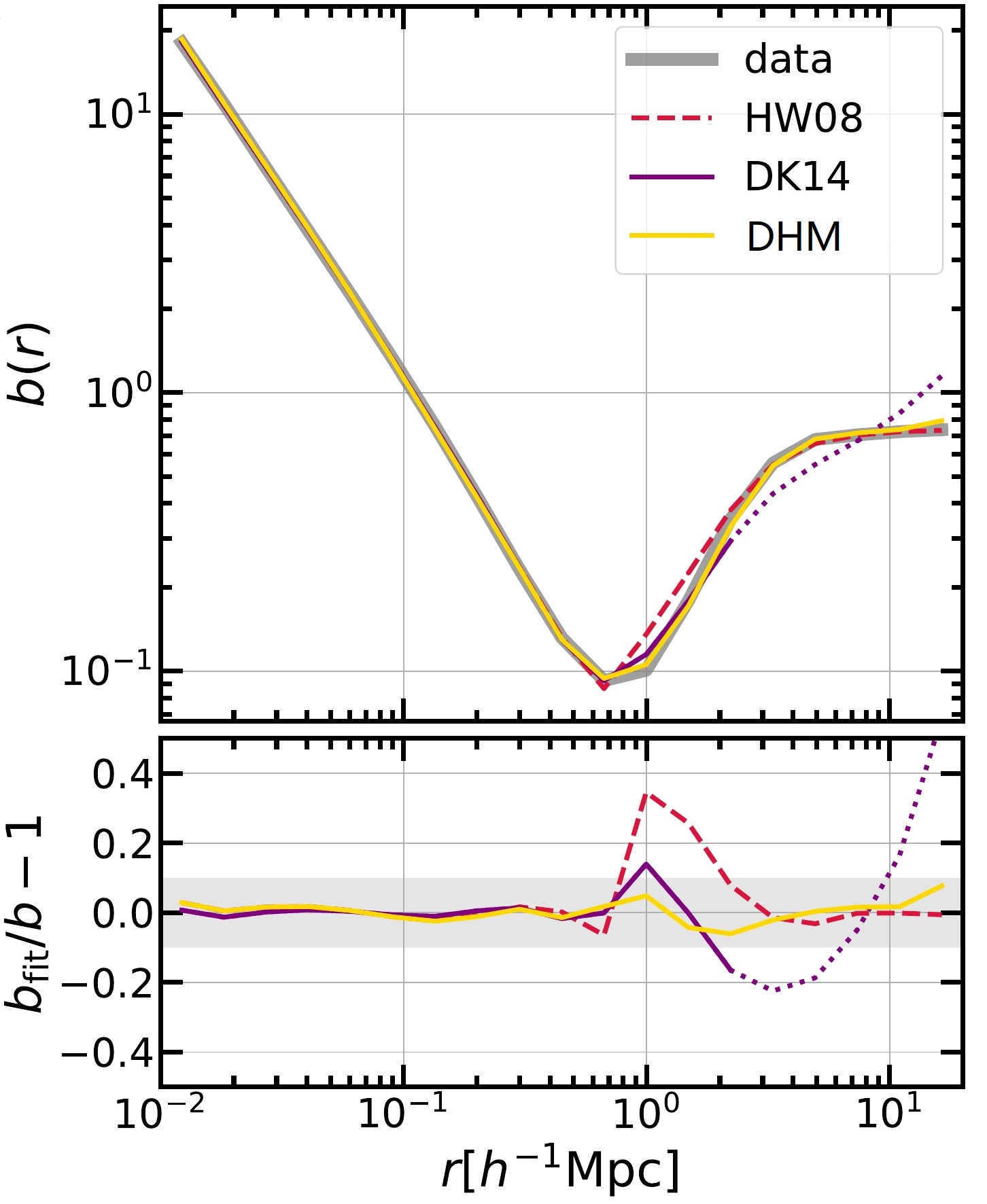}%
    \includegraphics[width=0.5\linewidth]{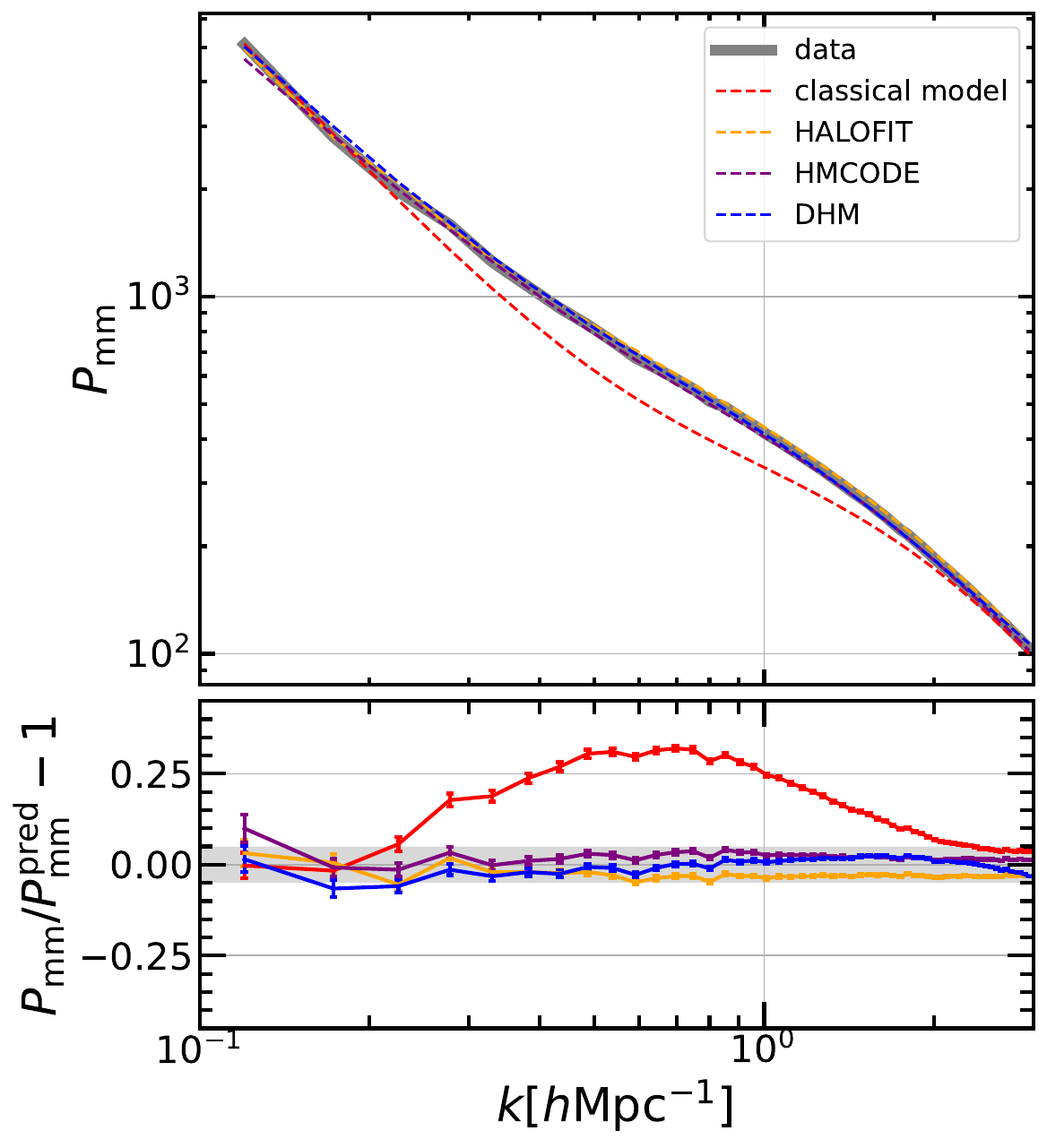}
    \caption{The performance of the depletion halo model (DHM) in predicting the halo and matter clustering. \textbf{Left:} The halo bias profile around a sample of $M_{\rm vir}\approx10^{12}\msunh$ halos. In the upper panel, the gray solid curve shows the simulation measurement, and the colored curves show predictions from the DHM, a classical halo model~\citep[HW08;][]{HayashiWhite2008} and the extended profile fitting of DK14 \citep{DK14}. The dotted part of DK14 represents extrapolation beyond their original fitting range. \textbf{Right:} The matter-matter power spectrum. The simulation measurements are compared against predictions from the DHM, a virial-radius based classical halo model, as well as \textsc{HaloFit} and \textsc{HMCode}. The bottom panels show the residuals of the models. The physical DHM achieves percent level accuracy in predicting both the halo-matter clustering and the matter-matter clustering across scales. See \citet{ZhouHan23} and \citet{ZhouHan25} for the original figures and further details.}
    \label{fig:DHM}
\end{figure}

Based on these ingredients, \citet{ZhouHan23,ZhouHan25} constructed a complete halo model at $z=0$ based on the depletion radius as a halo boundary. This depletion halo model (DHM) is essentially a ``vanilla'' first-principle model according to Equation~\ref{eq:matter_field}. While previous works struggle to model the halo-matter correlation on the inter-halo scale, the DHM can achieve $\sim 5\%$ accuracy in predicting the halo-matter and matter-matter clustering across linear and nonlinear scales without any fine-tuning or gluing of the model components (see Figure~\ref{fig:DHM}).

The accuracy of the DHM is comparable to those achievable by state-of-the-art halo-based packages such as \textsc{HaloFit}\citep{HaloFit03,HaloFit12} and \textsc{HMCode}\citep{HMCODE15,HMCODE16,HMCODE21}. However, it should be noted that the latter two are highly optimized fitting packages for predicting the matter power spectrum, which do not explicitly account for or necessarily respect the true distribution of an actual halo population. By contrast, the DHM corresponds explicitly to a population of physically identifiable halos, and is capable of outputting the clustering of both the matter and the halo fields. 

The decomposition of a halo into an orbiting and an infalling component can also help to improve the modeling of the matter distribution around halos, as both components can be described by parametric fitting functions~\citep{DiemerDecompose2, DiemerDecompose3, DynHaloModel,DynHaloProf}. Complete models for describing various large scale structure statistics using these halo definitions, including the total matter correlation function, remain to be developed. 

\subsubsection{Cosmological application}
The importance of the halo boundaries in modelling structure formation implies that they can also play a role in constraining cosmology using observations. \citet{Mpetha24} showed that the infall region, characterized by the dynamical boundaries such as the splashback and depletion radii, contains complementary cosmological information to those encoded in cluster abundance and cosmic shear~\citep[see also,][]{Haggar24}. Observational measurements on the halo profile or the boundaries in this region can help to break the $\sigma_8$-$\Omega_{\rm m}$ degeneracy associated with those late-time cosmological probes~\citep{Mpetha25}. 

Beyond the standard cosmological model, \citet{Adhikari18} showed that the splashback radius is sensitive to the equation of state of dark energy, due to its dependence on the expansion history of the universe. Besides, in theories of modified gravity involving a screening mechanism, the transition scale between screened and unscreened regions occurs in the halo outskirts, leaving detectable impacts on the location of the splashback radius~\citep{Adhikari14,Contigiani19}.

Further out, the turnaround radius can be used as another independent cosmological probe, as the turnaround density is a constant at a given redshift and depends on the cosmology~\citep{Korkidis20,Pavlidou20}, and it may be possible to measure the turnaround mass and radius in observations~\citep{Korkidis24, Korkidis25}. The upper limit on the turnaround radius, i.e., the maximum turnaround radius (Equation~\eqref{eq:rta_max}), has also been proposed as a potential probe for dark energy~\citep{Pavlidou14,Benisty25} although deviations from the spherical symmetry can result in deviations from the spherical collapse prediction~\citep{LeeYepes16,Bhattacharya21,Korkidis23,Benisty24a}. More generally, some works have made use of the detailed infall orbit and the galactic rotation curves to place constraints on dark energy (e.g., \cite{Benisty23,Benisty24b}).

\section{Future prospects}
While recent years have witnessed significant developments on the multi-layer structure of dark matter halo, there are still plenty of questions to explore ahead. Below we discuss a few aspects for future developments. With all these potential future progresses, it may be expected that a physics-based halo model will eventually provide us a much simpler yet comprehensive and accurate understanding of the structure formation process in the Universe. Such first-principle understandings offer vital competitions to, and could also be used complementarily to, effective and technical approaches such as 
the effective field theory~\citep[e.g.,][]{EFTRev,Desjacques18,Shuren25} or 
cosmological emulators~\citep[e.g.,][]{Zhai19,DarkQuest,CSSTemu} in the era of precision cosmology.

\subsection{Theoretical understanding}
Although the self-similar spherical collapse model can give an overall good prediction of the splashback radius, many simulation works have clearly pointed out the anisotropic behaviour of halo accretion and the corresponding boundary features. At the same time, analytical modelling of the anisotropic collapse remains poorly explored, not to say when putting halos into the larger scale environment which displays complex morphologies including filaments and sheets. In addition, in the ideal self-similar spherical collapse model, some of the outer boundaries, including the splashback radius, the depletion radius, and the edge radius, are identical, while in simulated halos the depletion radius is much larger than the splashback radius, and there is a large spread in the individual splashback radius. The relations between the shock radius and the splashback radius are also found to be complex and different works have not reached good concensus. A more advanced model should be able to predict the differences in these boundaries. Deviation from spherical symmetry could be a major contribution to the differences, in addition to the dispersion in particle orbital parameters and the influence of the external tidal field. For example, when the bulk acceleration of a halo due to the tidal field is accounted for via the so-called boosted potential~\citep{Stucker21}, it was found that the infalling particles and those on their first splashback orbit become unbound~\citep{strawberry}, in contrast to expectations from the spherical collapse models.

Going beyond the modelling of the boundaries, a more complete description of the halo structure resides in the phase space distribution of the surrounding particles. If halo collapse is fully described by the self-similar spherical collapse model, then the phase space structure of a halo is already fully specified theoretically.
However, the self-similar solution is found to be insufficient in predicting the properties of each individual stream in the multi-stream region (see discussions in section~\ref{sec:phasespaceboundary}), even after introducing non-zero angular momentum. 

Another major challenge to the theoretical understanding of halo structure and boundary is the effect of halo merger. The classical models of halo collapse have all assumed that the mass accretion is a smooth process, which remains a good approximation for the accretion of very low mass objects. Major mergers, however, differ significantly from smooth accretion in at least two aspects. First, a single merger brings in a large and localized mass contribution which takes on very little symmetry in its mass distribution around the central halo. The satellite mass distribution is also highly distorted during the interaction with the host halo. Secondly, major mergers are relatively rare and discrete, and cannot be modelled as a continuous process. Meanwhile, they can play important roles in determining the profile and boundary of a halo, and significantly influence the orbits of halo particles~\citep[e.g.,][]{Genina23,Lebeau24,Jiale}.
In particular, previous studies have also shown that mergers can drive merger-accelerated shocks in the gas and are responsible for the outermost accretion shock around a halo~\citep{Congyao20,Sen26}. 
A successful model of the merger feature could open up a new window for studying the merger-based accretion history of the halo.

\subsection{Observational measurements}
On the observational side, while there have been a number of measurements of the outer boundaries including the splashback radius and depletion radius for galaxy clusters and the Milky Way, the uncertainties in these measurements are still large and await improvements with more data. In particular, the major advantage of outer boundaries as halo growth rate diagnostics has not been fully exploited, due to the difficulty to accurately measure the outer boundaries for individual halos. Alternatively, one may stack halos according to observables that are expected to correlate with the halo growth rate, for example the galaxy concentration, color, or magnitude gap of the system. However, measurements along this line are still in tension with theoretical expectations~\citep[e.g.,][]{More16,Fong22}, which require careful understanding of systematics in the observational analysis and theoretical interpretation.

Despite the observational challenges, the outer boundaries could provide new proxies for the virial radius through scaling relations~\citep[e.g.,][]{Contigiani23,Gabriel-Silva25}, as the outer boundary may be measured directly from boundary features in the tracer distribution such as the steepening in the galaxy profile, while the virial radius lacks such point estimates. As an example, \cite{VallesPerez24} found that the shock radius is tightly correlated with the mass of clusters (within $2R_{\rm vir}$) and the Mach number, forming a plane in the three-dimensional parameter space.

\subsection{Applications in structure and galaxy formation}
The ultimate goal of studying the formation and structure of the halo is to improve our understanding and modelling of structure formation. The new boundaries are just starting to demonstrate their advantages in the modelling of large scale structure~\citep{Garcia21,ZhouHan23,ZhouHan25,DiemerDecompose3, DynHaloModel}. Despite this, much work is still needed to investigate the cosmological dependence of the models, and to build first principle analytical predictions for the model components as has been done in the EPS formalism for classical halos.

It is worth pointing out that it is not necessary to assume all the mass is locked in halos. For example, some works also account for a smoothly distributed matter component in addition to that in halos~\citep{ADM1, ZhouHan23}. A recent work extends the halo model to also account for lower-level structures including filaments, sheets and non-collapsed material~\citep{WebHM}. It could be expected that such a more general decomposition may serve as a more natural framework for describing the anisotropic large scale structure, due to asymmetries in the filament and sheet structures. Nevertheless, it remains important to understand and model the structure and boundaries of halos in this generalized framework, as well as their distributions in different environments, to provide the necessary small scale information not encoded in the lower level objects.

The application of the new halo boundaries to galaxy formation modelling is currently a substantially under-explored field. Existing analytical models of galaxy formation physics are primarily built upon the classical virial halo definition, with galaxies separated into central and satellites according to the virial radius, and the heating and cooling of the halo gas modelled within the virial radius~\citep[e.g.,][]{BensonRev}. As the outer boundary features are also present in the gas, star and galaxy components, it is natural to expect that the new boundaries play a role in the formation and evolution of galaxies. Observationally, \citet{Baxter17} showed that the red galaxy fraction shows a steepening around the splashback radius, consistent with an ansatz that the red galaxies correspond to an orbiting population while the blue ones are largely infalling. \citet{Zheng24} also found that the quenched fraction of galaxies show a transition around a scale consistent with the depletion radius. A new classification of central and satellite galaxies according to these new boundaries could also have significant implications for statistical studies connecting galaxy properties to halo properties, as the backsplash objects can be naturally identified as subhalos under the dynamical boundary~\citep[e.g.,][]{DiemerFlyby}. 

\subsection{Machine learning}
On the methodology side, machine learning can also play an important role for providing key insights into the physics of halo formation and structure, for example by identifying major components making up the universal density profile of a halo~\citep{LucieSmith22}, or by learning the key connections between the final halo mass profile and the initial condition or the accretion history~\citep{LucieSmith22b}. Encouragingly, these works can rediscover the component related to the infalling material or the outer edge of the halo without prior knowledge, supporting the significance of an outer component outside the virialized part. Machine learning also provides a flexible and non-parametric way to classify halo particles into different components~\citep[e.g.,][]{Narayan25}, or to tag galaxies into distinct classes such as central versus satellite or infalling versus orbiting ones~\citep[e.g.,][]{Farid23,Bowden25}. Some of these analyses can be done using only observable properties as input, thus laying the path to observationally identify galaxies according to the dynamical boundaries.

\begin{acknowledgements}
We thank Yifeng Zhou for providing a customized plot of their DHM result shown in Figure~\ref{fig:DHM}, and Xun Shi, Congyao Zhang, Benedikt Diemer, Daisuke Nagai, Tamara Richardson and Yanchuan Cai for helpful discussions and comments. This work is funded by the National Natural Science Foundation of China (No. 12595312), National Key R \& D Program of China (2023YFA1607800, 2023YFA1607801), China Manned Space Program (CMS-CSST-2025-A04), and Office of Science and Technology, Shanghai Municipal Government (grant Nos. 24DX1400100, ZJ2023-ZD-001). The computation of this work is carried out on the \textsc{Gravity} supercomputer at the Department of Astronomy, Shanghai Jiao Tong University.
\end{acknowledgements}

\appendix                  

\section{Practical equations for evaluating the self-similar model}\label{app:practical_eqs}
\subsection{Density, mass flow rate and velocity profiles}
The density profile can be found as
\begin{align}
\rho(r,t)&=\frac{\partial M(r,t)}{4\pi r^2\partial r}\\
&=\frac{M_{\rm ta}(t)}{4\pi r^2 R_{\rm ta}}\mathcal{M}'(x)\label{eq:rho_FG}\\
&=\rho_{\rm ta}(t)\frac{\mathcal{M}'(x)}{3x^2},\label{eq:density_prof}
\end{align} where $x\equiv r/R_{\rm ta}$ and $\mathcal{M}'(x)=\ud \mathcal{M}(x)/\ud x$.


The average radial velocity is then obtained as
\begin{align}
\bar{v}(r,t)&=\frac{\mu(r,t)}{4\pi r^2\rho(r,t)}\label{eq:vprof_def}\\
&=\frac{R_{\rm ta}}{t}[(\frac{2}{3}+\frac{2}{9\epsilon})x-\frac{2}{3\epsilon}\frac{\mathcal{M}(x)}{\mathcal{M}'(x)}]\\
&=\frac{R_{\rm ta}}{t}\frac{2}{3\epsilon}[(\epsilon+\frac{1}{3})-\frac{1}{\gamma(x)}]x\label{eq:vel_prof}
\end{align}

The derivative of $\mathcal{M}(x)$ can be evaluated from numerical derivative of $\mathcal{M}(x)$, or computed from the orbit through
\begin{align}
\mathcal{M}'(x)&=\frac{2}{3\epsilon}\int_1^{\infty} \frac{\ud \tau}{\tau^{1+2/(3\epsilon)}}\delta(\frac{\lambda(\tau)}{\Lambda(\tau)}-x)\mathrm{sign}(\frac{\ud x}{\ud \tau})\\
&=\frac{2}{3\epsilon}\sum_{\frac{\lambda(\tau)}{\Lambda(\tau)}=x}\frac{1}{\tau^{1+2/(3\epsilon)}|\frac{\lambda'}{\lambda}-\frac{\Lambda'}{\Lambda}|x}\label{eq:mass_deriv}
\end{align} where the summation is in the domain $\tau>1$.

According to the continuity equation, the mass flow rate $\mu\equiv -\dfrac{\partial M}{\partial t}$ can be alternatively found as $\mu=4\pi r^2 \rho v$. Utilizing this, the mass flow rate can be similarly derived as
\begin{align}
\mu(r)&=\sum_{\frac{\lambda(\tau)}{\Lambda(\tau)}=r/R_{\rm ta}} v(\tau) |\frac{\ud M_\ri(\tau)}{\ud r(\tau)}|\\
&=\frac{M_{\rm ta}}{t} \frac{2}{3\epsilon}\sum_{\frac{\lambda(\tau)}{\Lambda(\tau)}=x} \frac{\lambda'}{\lambda}\frac{1}{\tau^{2/(3\epsilon)}|\frac{\lambda'}{\lambda}-\frac{\Lambda'}{\Lambda}|}.\label{eq:mu_eval}
\end{align} The velocity profile is then found by Equation~\eqref{eq:vprof_def}.

\subsection{Pre-shell crossing profile and splashback/depletion density}
The evolution of a shell in the single-stream regime, i.e., before reaching the first caustic ($x>x_{\rm sp}$), follows the classical spherical collapse model without shell crossing. Even though Equation~\eqref{eq:SS_mass_prof_sum} is derived for $\tau\geq 1$, it is also applicable to the pre-turnaround phase with $\tau<1$ and $x>1$. In this coherent collapse phase, each radius is contributed by a single mass shell (i.e., $x(\tau)=x_0$ has only one solution), the mass enclosed by the orbit $x(\tau|M_\ri)$ simplifies to $\mathcal{M}(x(M_\ri))=M_\ri/M_{\rm ta}$, and the equation of motion reduces to the classical spherical collapse version. Now the orbital evolution is (c.f., Equations~\eqref{eq:sc_sol_r} and ~\eqref{eq:sc_sol_t})
\begin{align}
	\lambda&=\frac{1-\cos\phi}{2},\\
	\tau&=\frac{\phi-\sin\phi}{\pi}.
\end{align} The corresponding mass profile is
\begin{equation}
	\mathcal{M}(x)=\tau^{-2/(3\epsilon)},
\end{equation} where $x=\lambda/\Lambda=\lambda/\tau^{2/3+2/(9\epsilon)}$.

The density, velocity and MFR profiles can be obtained with previous equations \eqref{eq:mass_deriv}, \eqref{eq:density_prof}, \eqref{eq:vel_prof} and \eqref{eq:mu_eval} in the single stream regime, using
\begin{align}
	\frac{\lambda'}{\lambda}&=\frac{\pi\sin\phi}{(1-\cos\phi)^2},\\
	\frac{\Lambda'}{\Lambda}&=(\frac{2}{3}+\frac{2}{9\epsilon})\frac{1}{\tau}.
\end{align}


Solving these equations to find the time of the first shell-crossing when $x(\tau_{\rm id})=x_{\rm id}$,\footnote{While $x_{\rm id}=x_{\rm sp}$ in this model, we use $\tau_{\rm id}$ and $x_{\rm id}$ here to indicate the first time when the radius drops to the splashback/depletion radius, to avoid confusion with the time when the shell splashback. } the enclosed density inside the splashback/depletion radius , can be evaluated as
\begin{align}
	\rho_{\rm id}&=\rho_{\rm ta} \frac{\mathcal{M}(x_{\rm id})}{x_{\rm id}^3},\\
	&=\rho_{\rm ta} \frac{\tau_{\rm id}^2}{\lambda_{\rm id}^3}.
\end{align} This result depends on the initial perturbation slope, $\epsilon$, and hence on the mass accretion rate. For $\epsilon=0.3,0.8$, respectively, it can be found that 
$\tau_{\rm id}=1.84,1.76$, $\lambda_{\rm id}=0.45, 0.58$, $r_{\rm id}/R_{\rm ta}=0.19,0.34$ and $\rho_{\rm id}/\rho_{\rm ta}=37,16$. As $\rho_{\rm ta}\simeq 5.55\bar{\rho}$, this yields $\rho_{\rm id}/\bar{\rho}\simeq 205,88$ in the two cases.

\section{The \spheric\space package}\label{app:spheric}

\spheric is a compact Python implementation of the two spherical-collapse
models used in this review.\footnote{Version~0.1.2, available at
\url{https://github.com/Kambrian/SpheriC}.} It has been used to create Figures~\ref{fig:sc_sol} and \ref{fig:ss_solution}. 
Installation and usage instructions along with algorithmic implementation details are found in the package. Below we provide an overview of the design and scope of the package.

The package exports two classes.  \texttt{SCSolver} handles monolithic top-hat
collapse in a $\Lambda$CDM cosmology, and \texttt{ReducedOrbit} handles
self-consistent, self-similar secondary infall in an Einstein--de~Sitter
universe.

\texttt{SCSolver} accepts the present-day density parameters $\Omega_{\rm m,0}$
and $\Omega_{\Lambda,0}$, and returns the cosmic age, turnaround overdensity, and
virial overdensities relative to the critical and mean densities at a given scale
factor.  It also provides the pre-turnaround time--radius mapping for an
individual shell (the post-turnaround mapping follows by symmetry).  It
integrates the dimensionless perturbation expansion equation,
Equation~\eqref{eq:aprate_reduce}, and derives turnaround and virial quantities
from the standard timing and energy relations of monolithic collapse.

\texttt{ReducedOrbit} implements the Fillmore--Goldreich self-similar model.
The fundamental challenge is that the equation of motion for a shell
(Equation~\eqref{eq:SS_EoM}) depends on the mass
profile $\mathcal{M}(x)$, which is itself determined by the orbits of all
shells.  The code resolves this through fixed-point iteration, starting from a
trial $\mathcal{M}(x)$, integrating the shell orbit, and reconstructing the
profile by counting all radius-crossing events via
Equation~\eqref{eq:SS_mass_prof_sum}, then iterating until convergence.  A
separate difficulty is that the radial acceleration diverges as $\lambda\to0$
(Equation~\eqref{eq:SS_EoM}), so the orbit integration must be regularized near
the centre.  The package offers two numerical regularization strategies.  The
first retains a small angular momentum ($j>0$), whose centrifugal term
$j^2/\lambda^3$ holds the orbit away from the origin
(Equation~\eqref{eq:SS_EoM_j}), and in the limit $j\to0$ this approximates the
radial solution.  The second uses a reflecting boundary at a negligible radius
for a strictly radial orbit ($j=0$), treating the potential as static within it.
Both approaches require convergence verification with respect to the
regularization scale.  The former is integrated with a standard
Runge--Kutta--Fehlberg 7(8) scheme, the latter with a reflecting-boundary
variant.  After \texttt{solve()}, the methods \texttt{GetMass},
\texttt{GetDensity}, \texttt{GetMFR}, and \texttt{GetVel} return the
dimensionless profiles (Equations~\eqref{eq:SS_mass_prof_sum}--\eqref{eq:vprof_def}),
which reproduce the calculations of Section~\ref{sec:selfsimilar-sol}.

The method \texttt{locate\_bounds()} identifies the pericentres, apocentres, and caustics in the orbits.
\texttt{locate\_depletion()} computes the properties of the first (outermost) caustic, which in the idealized self-similar model coincides with both the splashback and depletion radii, as derived in Section~\ref{sec:ss_solution}.  The radial cutoff $r_{\rm conv}$, timestep limits, angular momentum $j$, and mass-profile iteration
tolerance are all user-adjustable with sensible defaults, and convergence checks
are recommended only for high-precision applications.

\bibliographystyle{raa}
\bibliography{bibtex}

\label{lastpage}

\end{document}